\documentstyle[11pt,amssymb,epsf]{article}
\def\crampest{\medmuskip = 1mu plus 1mu minus 1mu}
\def\uncramp{\medmuskip = 4mu plus 2mu minus 4mu}
\def\ben{\begin{equation}}
\def\een{\end{equation}}

\let\a=\alpha    
    
  \let\n=\nu

\let\C=\Chi 
  \let\re=\ref
  
\def\nn{\nonumber} \def\bd{\begin{document}} \def\ed{\end{document}}
\def\ds{\documentstyle} \let\fr=\frac \let\bl=\bigl \let\br=\bigr
\let\Br=\Bigr \let\Bl=\Bigl
\let\bm=\bibitem
\let\na=\nabla
\let\pa=\partial \let\ov=\overline
\newcommand{\be}{\begin{equation}}
\newcommand{\ee}{\end{equation}}
\def\ba{\begin{array}}
\def\ea{\end{array}}
\def\ft#1#2{{\textstyle{{\scriptstyle #1}\over {\scriptstyle #2}}}}
\def\fft#1#2{{#1 \over #2}}
\def\del{\partial}
\def\vp{\varphi}
\def\sst#1{{\scriptscriptstyle #1}}
\def\oneone{\rlap 1\mkern4mu{\rm l}}
\def\td{\tilde}
\def\wtd{\widetilde}
\def\ie{\rm i.e.\ }
\def\dalemb#1#2{{\vbox{\hrule height .#2pt
        \hbox{\vrule width.#2pt height#1pt \kern#1pt
                \vrule width.#2pt}
        \hrule height.#2pt}}}
\def\square{\mathord{\dalemb{6.8}{7}\hbox{\hskip1pt}}}
\newcommand{\ho}[1]{$\, ^{#1}$}
\newcommand{\hoch}[1]{$\, ^{#1}$}
\newcommand{\bea}{\begin{eqnarray}}
\newcommand{\eea}{\end{eqnarray}}
\newcommand{\ra}{\rightarrow}
\newcommand{\lra}{\longrightarrow}
\newcommand{\Lra}{\Leftrightarrow}
\newcommand{\ap}{\alpha^\prime}
\newcommand{\bp}{\tilde \beta^\prime}
\newcommand{\tr}{{\rm tr} }
\newcommand{\Tr}{{\rm Tr} }
\def\0{{\sst{(0)}}}
\def\1{{\sst{(1)}}}
\def\2{{\sst{(2)}}}
\def\3{{\sst{(3)}}}
\def\4{{\sst{(4)}}}
\def\5{{\sst{(5)}}}
\def\6{{\sst{(6)}}}
\def\7{{\sst{(7)}}}
\def\8{{\sst{(8)}}}
\def\n{{\sst{(n)}}}
\def\cA{{{\cal A}}}
\def\cF{{{\cal F}}}
\def\tV{\widetilde V}
\def\tW{\widetilde W}
\def\tH{\widetilde H}
\def\tE{\widetilde E}
\def\tF{\widetilde F}
\def\tA{\widetilde A}
\def\im{{{\rm i}}}
\def\tY{{{\wtd Y}}}
\def\ep{{\epsilon}}
\def\vep{{\varepsilon}}
\def\R{\rlap{\rm I}\mkern3mu{\rm R}}
\def\bD{{{\bar D}}}

\def\R{\rlap{\rm I}\mkern3mu{\rm R}}
\def\bD{{{\bar D}}}
\def\R{{{\Bbb R}}}
\def\C{{{\Bbb C}}}
\def\H{{{\Bbb H}}}
\def\CP{{{\Bbb C}{\Bbb P}}}
\def\RP{{{\Bbb R}{\Bbb P}}}
\def\Z{{{\Bbb Z}}}
\def\bA{{{\Bbb A}}}
\def\bB{{{\Bbb B}}}
\def\bC{{{\Bbb C}}}
\def\bR{{{\Bbb R}}}
\def\bD{{{\Bbb D}}}
\def\bE{{{\Bbb E}}}
\def\bZ{{{\Bbb Z}}}
\def\Re{{{\frak{Re}}}}
\def\Im{{{\frak{Im}}}}
\def\cosec{{\,\hbox{cosec}\,}}
\def\Gm{{\Gamma_{\!\! -}}}
\def\Gp{{\Gamma_{\!\! +}}}
\def\stan{{standard }}
\def\nonstan{{supernumerary }}

\def\cosech{{\hbox{cosech}}}

\def\etcyc{{\hbox{and cyclic}}}
\def\btheta{{\bar\theta}}

\newcommand{\tamphys}{\it Center for Theoretical Physics,
Texas A\&M University, College Station, TX 77843, USA}
\newcommand{\umich}{\it Michigan Center for Theoretical Physics,
University of Michigan\\ Ann Arbor, MI 48109, USA}
\newcommand{\upenn}{\it Department of Physics and Astronomy,\\
University of Pennsylvania, Philadelphia,  PA 19104, USA}
\newcommand{\SISSA}{\it  SISSA-ISAS and INFN, Sezione di Trieste\\
Via Beirut 2-4, I-34013, Trieste, Italy}

\newcommand{\mitchell}{\it George P. and Cynthia W. 
Mitchell Institute for Fundamental Physics,\\
Texas A\&M University, College Station, TX 77843-4242, USA}

\newcommand{\newton}{\it Isaac Newton Institute for Mathematical Sciences,\\
20 Clarkson Road,  University of Cambridge,
Cambridge CB3 0EH, UK}

\newcommand{\ihp}{\it Institut Henri Poincar\'e\\
  11 rue Pierre et Marie Curie, F 75231 Paris Cedex 05}

\newcommand{\rutg} {\it Department of Physics and Astronomy,\\
Rutgers University, Piscataway, NJ 08855, USA } 

\newcommand{\damtp}{\it DAMTP, Centre for Mathematical Sciences,
 Cambridge University\\  Wilberforce Road, Cambridge CB3 OWA, UK}
\newcommand{\itp}{\it Institute for Theoretical Physics, University of
California\\ Santa Barbara, CA 93106, USA}

\newcommand{\auth}{M. Cveti\v{c}\hoch{\dagger 1},
H. L\"u\hoch{\ddagger 2},
C.N. Pope\hoch{\ddagger 2} and K.S. Stelle\hoch{\star 3} }

\begin{document}
\begin{flushright}
\hfill{RUNHETC-2002-34\ \ \
%CTP TAMU-20/02 \ 
MIFP-02-01\ \ \
Imperial/TP/01-02/27\ \ \
UPR-1013-T}
\\
\hfill{
\bf hep-th/0209193}
\end{flushright} 

\begin{center}  

{\large {\bf Linearly-realised Worldsheet Supersymmetry in pp-wave 
 Backgrounds}}   

\vspace{15pt}

\auth

\vspace{7pt}
{\hoch{\dagger}\rutg}

\vspace{7pt}
{\hoch{\ddagger}\mitchell}

\vspace{7pt}
{\hoch{\star}{\it The Blackett Laboratory, Imperial College, Prince
Consort Road, London SW7 2BZ. }}

\underline{ABSTRACT}
\end{center}  

   We study the linearly-realised worldsheet supersymmetries in the
``massive'' type II light-cone actions for pp-wave backgrounds.  The
pp-waves have have $16+N_{\rm sup}$ Killing spinors, comprising 16
``standard'' Killing spinors that occur in any wave background, plus
$N_{\rm sup}$ ``supernumerary'' Killing spinors ($0\le N_{\rm sup} \le
16$) that occur only for special backgrounds.  We show that only the
supernumerary Killing spinors give rise to linearly-realised
worldsheet supersymmetries after light-cone gauge fixing, while the 16
standard Killing spinors describe only non-linearly realised
inhomogeneous symmetries.  We also study the type II actions in the
physical gauge, and we show that although in this case the actions are
not free, there are now linearly-realised supersymmetries coming both
from the standard and the supernumerary Killing spinors. In the
physical gauge, there are no mass terms for any worldsheet degrees of
freedom, so the masses appearing in the light-cone gauge may be viewed
as gauge artefacts. We obtain type IIA and IIB supergravity solutions
describing solitonic strings in pp-wave backgrounds, and show how
these are related to the physical-gauge fundamental string actions.
We study the supersymmetries of these solutions, and find examples
with various numbers of Killing spinors, including total numbers that
are odd.

{\vfill\leftline{}\vfill
\vskip 10pt \footnoterule
{\footnotesize \hoch{1} 
On sabbatical leave from the University of Pennsylvania.
Research supported in part by DOE grant
\phantom{DOEF}DOE-FG02-95ER40893, 
NATO linkage  grant No. 97061 and  Class of 1965 Endowed Term Chair.
\vskip  -12pt} \vskip   14pt
{\footnotesize \hoch{2}
Research supported in part by DOE grant DE-FG03-95ER40917
\vskip -12pt} \vskip 14pt
{\footnotesize \hoch{3}
Research supported in part by the EC under RTN
contract HPRN-CT-2000-00131.
\vskip -12pt}  \vskip  14pt
}

\pagebreak
\setcounter{page}{1}

\tableofcontents
\addtocontents{toc}{\protect\setcounter{tocdepth}{2}}
\newpage

\section{Introduction}

    The pp-wave configuration \cite{kg,blafighulpap,blafighulpap2}
that arises as a Penrose limit \cite{penrose} of the AdS$_5\times S^5$
solution in string theory gives rise to an exactly-solvable free
massive worldsheet-supersymmetric string action in the light-cone
gauge \cite{met,bermalnas}, which can provide insights into aspects of
the AdS/CFT correspondence \cite{bermalnas}.  More general pp-wave
solutions of a similar kind also exist
\cite{clppp,supernum,gahu,lupo}, which may or may not be obtainable
from any Penrose limit.  They again give rise to exactly-solvable free
massive string actions.  The Penrose limit of AdS$_5\times S^5$ is
maximally supersymmetric, with 32 Killing spinors.  Typically, the
other pp-wave solutions have less supersymmetry, although every
pp-wave admits at least 16 Killing spinors, regardless of the specific
details of its construction.  In fact the Killing spinors divide into
two categories, which in the terminology of \cite{clppp,supernum} are
the 16 ``standard'' Killing spinors that every pp-wave has, and the
possible further ``supernumerary'' Killing spinors, whose number
$N_{\rm sup}$ ($0\le N_{\rm sup}\le 16$) depends upon the details of
the solution.

    An important question arises concerning the supersymmetry of the
light-cone string action.  Prior to gauge fixing, the Green-Schwarz
action has a local kappa symmetry and a rigid spacetime supersymmetry,
with a number of parameters equal to the number of Killing spinors in
the target spacetime.  After imposing the light-cone gauge conditions,
the kappa symmetry and spacetime supersymmetry transmute into rigid
worldsheet supersymmetries of the string action.  Of particular
interest are the worldsheet supersymmetries that are linearly
realised, since they establish a pairing between the bosonic and
fermionic degrees of freedom, and they can imply relations between the
masses of the bosons and the fermions.

   It was observed in \cite{clppp,supernum} that in general, for a
pp-wave that has only the 16 standard Killing spinors, the mass terms
for the bosons and the fermions are unequal, and thus evidently there
could be no linearly-realised worldsheet supersymmetries.\footnote{Or
at least none that commute with the Hamiltonian.}  By contrast, it was
observed that if there are supernumerary Killing spinors that in
addition are independent of the $X^+$ coordinate (and hence are
independent of the worldsheet time coordinate in light-cone gauge,
and so commute with the Hamiltonian) then the boson and fermion
masses {\it are} related.  This led to the conjecture in
\cite{clppp,supernum} that it is only the supernumerary Killing
spinors that can give rise to linearly-realised worldsheet
supersymmetries in the string action in the light-cone gauge.
    
   One of the main purposes of the present paper is to prove this
conjecture, by showing that when one constructs the light-cone-gauge
string action in a pp-wave background, linearly-realised worldsheet
supersymmetries are indeed associated purely with the {\it
supernumerary} Killing spinors that arise in special pp-wave
backgrounds.  By contrast, we show that the 16 standard Killing
spinors do not give rise to any linearly-realised worldsheet
supersymmetries.  We prove these results by showing that the
projection conditions for linearly-realised worldsheet supersymmetries
that arise from fixing the kappa symmetry of the Green-Schwarz string
action in the light-cone gauge eliminate all the standard Killing
spinors in the pp-wave, but allow all the supernumerary Killing
spinors to remain.  This shows that it is precisely the supernumerary
Killing spinors that are associated with linearly-realised worldsheet
supersymmetries in the light-cone gauge.

   A different approach to the question of linearly-realised
worldsheet supersymmetry is to construct a classical supergravity
solution describing a solitonic string on a pp-wave background.  The
dynamics of the string on this background arise from considering the
Goldstone modes associated with the breaking of symmetries of the
pp-wave background when the string is present. We construct such
intersecting pp-wave/string solutions.  A puzzle is that these always
preserve some non-vanishing fraction of the supersymmetry, even in
cases where the pp-wave by itself would have no supernumerary
supersymmetries.  One expects to find, therefore, that the Goldstone
boson and fermion modes are still related by a linearly-realised
supersymmetry, associated with the residual unbroken supersymmetry of
the pp-wave/string solution, even in cases without supernumerary
supersymmetries.  This would then imply that the associated string
action, which describes the dynamics of the Goldstone modes, should
have linearly realised worldsheet supersymmetry regardless of whether
or not there exist supernumerary Killing spinors in the pure pp-wave
background.

   The resolution of this apparent discrepancy with the previous
light-cone discussion is that the solution with a string on a pp-wave
leads to Goldstone modes describing the string dynamics in the {\it
physical gauge} and not the light-cone gauge.  (In the physical gauge,
two of the ten target spacetime coordinates are set equal to the two
string worldsheet coordinates $(\tau,\sigma)$, rather than setting
$X^+=\tau$ as one does in light-cone gauge.)  We are therefore led to
re-examine the string action in the physical gauge, to see what
conclusions one then reaches about linearly-realised worldsheet
supersymmetries in a pp-wave background.

   We study this issue by starting from the Green-Schwarz string
action, and fixing the kappa symmetry in physical gauge, in order to
derive the residual linearly-realised worldsheet supersymmetry.  We
find that, unlike the analogous calculation in light-cone gauge, where
the projection conditions remove all 16 of the standard Killing
spinors of a generic pp-wave background, in physical gauge the
projection can instead preserve a certain fraction of the
standard Killing spinors.  It
also eliminates some of the supernumerary Killing spinors, while
preserving others.  Thus in the physical gauge, there is
linearly-realised worldsheet supersymmetry for {\it any} pp-wave
background, with enhanced worldsheet supersymmetry if there are also
supernumerary Killing spinors.

   In section 2 we consider the type IIA Green-Schwarz action, up to
terms of quadratic order in fermions, and we exhibit the kappa symmetry
and spacetime supersymmetry transformations.  We then derive the rigid
worldsheet supersymmetries that survive after imposing the light-cone
gauge conditions.  In section 3, we give an analogous discussion for the
type IIB Green-Schwarz action.  In section 4, we show how the
projection conditions for linearly-realised worldsheet supersymmetries
in the type IIA and type IIB light-cone actions are compatible with
those satisfied by the supernumerary Killing spinors, but they are
orthogonal to the projection conditions satisfied by the 16 standard
Killing spinors.  In section 5 we turn to an analysis of the physical
gauge, deriving the residual worldsheet supersymmetries after
gauge-fixing. The projection conditions on the residual
supersymmetries turn out to be compatible with standard as well as
supernumerary Killing spinors in the physical gauge.  In section 6, we
construct new supergravity solutions describing strings in pp-wave
backgrounds, and show how these provide realisations for string
actions in pp-wave backgrounds in the physical gauge. Various unusual
fractions of supersymmetry can be achieved in the string/pp-wave
solutions, including odd numbers of Killing spinors.

\section{Type IIA Green-Schwarz Action}

\subsection{Kappa symmetry and supersymmetry in type IIA}

   The explicit component form of the Green-Schwarz action for the
type IIA string on an arbitrary bosonic background, up to and
including terms quartic in the fermionic coordinates, was derived in
\cite{clps} This result was obtained by double dimensional reduction of
an explicit component form of the supermembrane action in $D=11$ that
was obtained in \cite{dwpp}.  The reduction to the type IIA string that was
performed in \cite{clps} was a component analogue of the superfield
reduction performed in \cite{duhoinst}.  The type IIA action found 
in \cite{clps} is 
%%%%%
\bea
{\cal L}_2 &=& -\ft12 \sqrt{-h}\, h^{ij}\, \del_i X^\mu\, \del_j
X^\nu\, g_{\mu\nu} + \ft12\ep^{ij}\, \del_i X^\mu\, \del_j X^\nu\,
A_{\mu\nu} \nn\\
&&- \im\, \btheta\, \beta^{ij}\,
\Gamma_\mu\, D_j\theta\, \del_i X^\mu +\ft{\im}8
\del_i X^\mu\, \del_j X^\nu\, \btheta\,\beta^{ij}\,\Gamma_{11}\,
\Gamma_\mu{}^{\rho\sigma}\, \theta\,
F_{\nu\rho \sigma}  \label{type2alag}\\
&&-\ft{\im}{16} \del_i X^\mu \del_j X^\nu e^{\phi}\, \btheta
\,\beta^{ij}\, \Big(\Gamma_{11}\,
\Gamma_\mu\, \Gamma^{\rho\sigma}\, \Gamma_\nu\, F_{\rho\sigma} + \ft1{12}
\Gamma_\mu\, \Gamma^{\rho\sigma\lambda\tau}\, \Gamma_\nu\,
F_{\rho\sigma\lambda\tau} \Big)\, \theta\,,\nn
\eea
%%%%%
where
%%%%%
\be
\beta^{ij} \equiv \sqrt{-h}\, h^{ij} -\ep^{ij}\, \Gamma_{11}\,,\qquad
D_i\theta \equiv  \del_i\theta + \ft14 \del_i X^\mu\,
\omega_\mu{}^{mn}\, \Gamma_{mn}\, \theta\,.\label{sdef}
\ee
%%%%%
The field strengths are given by
%%%%%
\be
F_\4 = dA_\3 - A_\1\wedge dA_\2\,,\qquad F_\3= dA_\2\,,\qquad
F_\2 = dA_\1\,.
\ee
%%%%%

   We can show that this action is invariant, up to the quadratic order in
$\theta$ to which we are working, under local kappa-symmetry
transformations given by
%%%%%
\be
\delta\, \theta = (1+\Gamma)\, \kappa\,,\qquad
\delta \, X^\mu = -\im\, \btheta\, \Gamma^\mu\, \delta\, \theta\,.
\ee
%%%%%
To the order we are working, the matrix $\Gamma$ is given by
%%%%%
\be
\Gamma= \fft1{2\sqrt{-h}}\, \ep^{ij}\, \del_i X^\mu\, \del_j X^\nu\, 
\Gamma_{\mu\nu}\, \Gamma_{11}\,.
\ee
%%%%%

   We can also show that the action is invariant, up to the relevant
order in $\theta$, under rigid supersymmetry transformations, given by
%%%%%
\be
\delta\, \theta=\ep\,,\qquad \delta\, X^\mu = \im\, \btheta\, \Gamma^\mu\, 
\ep\,,\label{2asusy}
\ee
%%%%%
where $\ep$ is a Killing spinor of the type IIA supergravity background.
Specifically, we find that under (\ref{2asusy}), the Lagrangian 
(\ref{type2alag}) varies to give, up to total derivatives,
%%%%%
\be
\delta \, {\cal L} = -2\im\, \del_i X^\mu\, \del_j X^\nu\, 
\btheta\, \beta^{ij}\, \Gamma_\mu\, {\cal D}_\nu\, \ep\,,
\ee
%%%%%
where ${\cal D}_\mu$ is the supercovariant derivative, given by
%%%%%
\be
{\cal D}_\mu = \nabla_\mu + \ft18 \Gamma_{11}\, \Gamma^{\rho\sigma}\, 
  F_{\mu\rho\sigma} - \ft1{16} e^\phi\, \Big(\Gamma_{11}\,  
\Gamma^{\rho\sigma}\, 
F_{\rho\sigma} - \ft1{12} \Gamma^{\rho\sigma\lambda\tau}\, 
F_{\rho\sigma\lambda\tau}\Big)\, \Gamma_\mu\,,
\ee
%%%%%
where $\nabla_\mu \equiv \del_\mu + \ft14 \omega_\mu^{mn}\, \Gamma_{mn}$.
In the string-frame metric that we are using here, the supersymmetry
transformation law for the gravitini is given by $\delta\, \psi_\mu = 
{\cal D}_\mu\, \ep$, and so the Green-Schwarz action is invariant under
(\ref{2asusy}) if $\ep$ is a Killing spinor.

   We may also define the derivative
%%%%%
\be 
\wtd{\cal D}_i = D_i + \del_i X^\mu\, \Big[\ft18 \Gamma_{11}\,
\Gamma^{\rho\sigma}\, F_{\mu\rho\sigma} - \ft1{16} e^\phi\, \Big(
\Gamma_{11}\, \Gamma^{\rho\sigma}\, F_{\rho\sigma} - \ft1{12}
\Gamma^{\rho\sigma\lambda\tau}\, F_{\rho\sigma\lambda\tau}\Big)\,
\Gamma_\mu\Big]\,.
\ee
%%%%%
When acting on a spinor function of $X^\mu$ such as $\ep$, this reduces to
$\wtd{\cal D}_i
\, \ep= \del_i X^\mu\, {\cal D}_\mu\, \ep$. In terms of $\wtd{\cal D}_i$, 
we can
write the  Green-Schwarz Lagrangian (\ref{type2alag}) as 
%%%%%
\be
{\cal L}_2 = -\ft12 \sqrt{-h}\, h^{ij}\, \del_i X^\mu\, \del_j
X^\nu\, g_{\mu\nu} + \ft12\ep^{ij}\, \del_i X^\mu\, \del_j X^\nu\,
A_{\mu\nu} -\im\, \del_i X^\mu\, \btheta\, \beta^{ij}\, \Gamma_\mu\, 
\wtd{\cal D}_j \theta\,.
\ee
%%%%%

\subsection{Type IIA worldsheet supersymmetry in the light-cone gauge}
\label{2alightcone}

    We begin with a review of how one imposes the light-cone gauge 
condition in the background of a pp-wave.  The covariant Green-Schwarz
action can be written as 
%%%%%
\be
{\cal L}= -\ft12 \sqrt{-h}\, h^{ij}\,\Pi_i^m\, \Pi_j^n\, \eta_{mn} 
-\ft12 \ep^{ij}\, \del_i Z^M\, \del_j Z^N\, A_{NM}\,,
\ee
%%%%%
where $\Pi_i^m = \del_i Z^M\, E^m_M$, $Z^M=(X^\mu, \theta_\a)$, and
the relevant supervielbeins are given, up to quadratic order in
fermion coordinates $\theta$, by
%%%%%
\bea
E_\mu^m &=& e^m_\mu + \ft{\im}4 \omega_\mu^{pq}\, \bar\theta \, \Gamma^m\,
\Gamma_{pq}\, \theta -\ft{\im}8 \bar\theta\, \Gamma_{11}\,
\Gamma^m\, \Gamma^{pq}\,
\theta\, F_{\mu pq} + \ft{\im}{16} e^\phi\, \bar\theta\, \Gamma_{11}\,
\Gamma^m\, \Gamma^{pq}\, \Gamma_\mu\, \theta\, F_{pq}\nn\\
&&
+\ft{\im}{192} e^\phi\, \bar\theta\, \Gamma^m\, \Gamma^{p_1\cdots p_4}\,
\Gamma_\mu\, \theta\, F_{p_1\cdots p_4}\,,\nn\\
E_\alpha^m &=& -\im\, (\bar\theta\Gamma^m)_\alpha\,.\label{2asuperv}
\eea
%%%%%
The worldsheet metric is given by
%%%%%
\be
h_{ij} = \Pi_i^m\, \Pi_j^n\, \eta_{mn}\,.\label{hfrompi}
\ee
%%%%%

    There is a local worldsheet (kappa) symmetry and a rigid spacetime 
supersymmetry,\footnote{Note that ``rigid'' here means that $\ep$ is
a Killing spinor in the ten-dimensional type IIA target spacetime, hence
determined by a finite number of constant parameters.}
with parameters $\kappa$ and
$\ep$ respectively,
%%%%%
\be
\delta\theta = (1+\Gamma)\, \kappa + \ep\,,\qquad
\delta X^\mu = -\im\, \bar\theta\, \Gamma^\mu\, (1+\Gamma)\, \kappa +\im\, 
 \bar\theta\,\Gamma^\mu\ep\,,\label{susytrans}
\ee
%%%%%
where 
%%%%%
\be
\Gamma = \fft{1}{2\sqrt{-h}}\, \ep^{ij}\, \Pi_i^m\, \Pi_j^n\, 
\Gamma_{mn}\, \Gamma_{11}\,.\label{2agamma}
\ee
%%%%%
(Note that we can write $\delta\, X^\mu = -\im\, \bar\theta\, \Gamma^\mu\,
\delta\theta + 2\im\, \bar\theta\, \Gamma^\mu\, \ep$.)
One can straightforwardly establish from the definitions that $\Gamma^2=1$
and $\tr \, \Gamma=0$, so $(1+\Gamma)$ projects out half the components of 
$\kappa$.
  
 In the light-cone gauge one imposes
%%%%%
\be
X^+=\tau\,,\qquad \Gamma^+\, \theta=0\,,\qquad 
\sqrt{-h}\, h^{ij} = \eta^{ij}\,.
\ee
%%%%%
From (\ref{2asuperv}) and the other definitions above, it therefore follows
that in cases where the bosonic vielbein satisfies 
%%%%%
\be
e^+_+=1\,,\qquad e^+_-=0\,,\qquad e^+_I=0\,,\label{boscon}
\ee
%%%%% 
(where we decompose the local Lorentz indices as $m=(+,-,I)$), we shall have
%%%%%
\be
\Pi_0^+ = 1\,,\qquad \Pi_1^+=0\,.
\ee
%%%%%
Note that (\ref{boscon}) is satisfied not only for a flat Minkowski
background spacetime, but also for the pp-wave solution.
Since the worldsheet metric is given by (\ref{hfrompi}), we also deduce
that in light-cone gauge we shall have
%%%%%
\bea
&&h_{00}=2 \Pi_0^- + \Pi_0^I\, \Pi_0^I\,,\qquad
h_{11}= \Pi_1^I\, \Pi_1^I\,,\qquad
h_{01} = \Pi_1^- + \Pi_0^I\, \Pi_1^I\,,\nn\\
&& h_{11}=-h_{00}=\sqrt{-h}\,, \qquad h_{01}=0\,.
\eea
%%%%%
Thus in particular we have
%%%%%
\be
 \Pi_0^- = -\ft12 (\Pi_0^I\, \Pi_0^I + \Pi_1^I\, \Pi_1^I)\,,\qquad
\Pi_1^- = -\Pi_0^I\, \Pi_1^I\,.
\ee
%%%%%

   It is useful at this stage to introduce the two matrices
%%%%%
\be
P\equiv \Pi_0^I\, \Gamma_I\qquad Q\equiv \Pi_1^I\, \Gamma_I\,,
\ee
%%%%%
in terms of which we have
%%%%%
\be
 \Pi_0^-  =-\ft12 (P^2+Q^2)\,,\qquad \Pi_1^- = -\ft12 (P\, Q+ Q\, P)\,.
\ee
%%%%%
The matrix $\Gamma$ defined in (\ref{2agamma}) can then be written in 
the light-cone gauge as 
%%%%%
\be
\Gamma=\fft1{\sqrt{-h}}\, \Big( \ft12 \Gamma_-\, \Gamma_+\, (P\, Q + Q\, P) 
+\ft12 \Gamma_-\, (P\, Q\, P - Q^3) + \Gamma_+\, Q - Q\, P\Big)\, 
\Gamma_{11}\,,\label{gammalc}
\ee
%%%%%%
and we can write $1/\sqrt{-h}$ as $Q^{-2}$.

   To derive the residual worldsheet supersymmetry that emerges from
light-cone gauge fixing of the kappa and rigid $\ep$ symmetries, we
first note that we must preserve the light-cone condition $\Gamma^+\,
\theta=0$ (\ie $\Gamma_-\, \theta=0$), implying that
%%%%%
\be
\delta(\Gamma_-\, \theta) = \Gamma_-(\kappa+\ep) + \Gamma_-\, \Gamma\, 
\kappa = 0\,.
\ee
%%%%%
From (\ref{gammalc}) this gives
%%%%%
\be
Q\, \Gamma_-\, (\kappa+\ep) + (\Gamma_-\, \Gamma_+ +\Gamma_-\, P)\, 
\Gamma_{11}\, \kappa=0\,.\label{lcpres}
\ee
%%%%%
This can be viewed as an equation determining 16 of the 32 components
of $\kappa$, in terms of $\ep$.  Thus the $\kappa$ transformation
is a compensating transformation that ensures the preservation of
the light-cone condition $\Gamma_-\, \theta=0$ under rigid supersymmetry
transformations. 
The residual symmetries will be given by $\delta(\Gamma_+\, \theta)=
\Gamma_+\, (\kappa+\ep) + \Gamma_+\, \Gamma\, \kappa$.  Using
(\ref{gammalc}) and (\ref{lcpres}), we find after simple algebra that
%%%%%
\be
\delta(\Gamma_+\, \theta) = \ft12( P + Q\, \Gamma_{11})\, \Gamma_+\, 
\Gamma_-\, \ep + \Gamma_+\, \ep\,.\label{residualsusy}
\ee
%%%%%
Note that $\kappa$ no longer appears; although one can solve for only
16 of the 32 components of $\kappa$ using (\ref{lcpres}), it is only
these components that enter in the expression for $\delta(\Gamma_+\,
\theta)$.

   The residual transformations (\ref{residualsusy}) comprise a
standard type of homogeneous worldsheet supersymmetry, together with
a shift symmetry.  It is clearest to see this by choosing a 
$32=16\otimes 2$ basis
where 
%%%%%
\be
\Gamma_+= \pmatrix{0& \sqrt2\cr 0& 0}\,,\quad
\Gamma_-= \pmatrix{0& 0\cr \sqrt2 & 0}\,,\quad
\Gamma_I = \pmatrix{\gamma_I & 0\cr 0& -\gamma_I}\,,\quad 
\Gamma_{11}=\pmatrix{\gamma_9&0\cr 0& -\gamma_9}\,,\label{162decomp}
\ee
%%%%%
where $\gamma_9=\prod_{I=1}^8 \gamma_I$, $\Gamma_\pm =(\Gamma_9\pm
\Gamma_0)/\sqrt2$.
One has
%%%%%
\be
\theta=\pmatrix{\theta_1 \cr \theta_2}\,,\qquad
\ep=\pmatrix{\ep_1 \cr \ep_2}\,,\qquad
\kappa=\pmatrix{\kappa_1 \cr \kappa_2}\,,
\ee
%%%%
and thus the gauge condition $\Gamma_-\, \theta=0$ implies $\theta_1=0$
while (\ref{residualsusy}) gives
%%%%%
\be
\delta\theta_2 = \ft1{\sqrt2}\, (p + q\, \gamma_9)\, \ep_1 + \ep_2\,.
\ee
%%%%%
Here 
%%%%%
\be
P= \pmatrix{p&0\cr 0 & -p}\,,\qquad 
Q= \pmatrix{q&0\cr 0 &-q}\,.
\ee
%%%%%
Since $p=\Pi_0^I\, \gamma_I$ and $q=\Pi_1^I\, \gamma_I$, we see that
$\ep_1$ describes a homogeneous (sometimes somewhat misleadingly 
called {\it linearly-realised}) supersymmetry, while $\ep_2$ describes 
an inhomogeneous shift symmetry.

   The corresponding supersymmetry transformation of the bosonic coordinates
in (\ref{susytrans}) then gives
%%%%%
\bea
\delta X^I &=& \im\, \bar\theta\, \Gamma^I\, [(1+\Gamma)\, \kappa -\ep]
+2\im\,  \bar\theta\, \Gamma^I\,\ep \,,\nn\\
&=& \ft{\im}{2} \, \bar\theta\, \Gamma^I\, \, \Gamma_+\, 
 [\Gamma_-\, (1+\Gamma)\, \kappa -\Gamma_-\, \ep] 
+2 \im\,  \bar\theta\, \Gamma^I\, \,  \ep\,,\nn\\
&=& 2\im\,  \bar\theta\, \Gamma^I\,\ep\,.
\eea
%%%%%
One also easily checks that $\delta X^+=0$, and so the light-cone gauge 
condition $X^+=\tau$ is preserved.  Thus in all we have
%%%%
\be
\delta \theta_2 = \ft1{\sqrt2}\, (p + q\, \gamma_9)\, \ep_1 + \ep_2\,,
\qquad
\delta X^I = 2\im\, \bar\theta_2\, \gamma^I\, \ep_1\,.
\ee
%%%%%

   We see in particular that the ``linearly-realised'' worldsheet 
supersymmetries corresponding to $\ep_1$ are associated with Killing
spinors $\ep$ that satisfy the projection condition
%%%%%
\be
\Gamma_+\, \ep=0\,.\label{linsusycon}
\ee
%%%%%
Note that the 16 ``ordinary'' Killing spinors in the pp-wave
background satisfy $\Gamma_-\, \ep=0$.  Thus we see that the condition
for having linearly-realised supersymmetry on the worldsheet precisely
conflicts with the condition satisfied by the ordinary 16 Killing
spinors in the pp-wave.  This means that none of the ordinary 16
Killing spinors gives rise to linearly-realised supersymmetry on the string
worldsheet in lightcone gauge.  By contrast, the supernumerary Killing
spinors in a pp-wave solution, which are themselves subject to the
opposite projection condition $\Gamma_+\, \ep=0$, are precisely 
compatible with the projection condition (\ref{linsusycon}) for 
linearly-realised worldsheet supersymmetries.

\section{Type IIB Green-Schwarz Action}

\subsection{Kappa symmetry and supersymmetry in type IIB}

    The explicit component form of the Green-Schwarz action for the
type IIB string on an arbitrary bosonic background was derived in
\cite{clps}, up to and including terms quadratic in the fermionic
coordinates.  The action was derived by explicitly implementing a
T-duality transformation of the type IIA Green-Schwarz action in 
\cite{clps}.  

    The type IIB Green-Schwarz action obtained in \cite{clps} was
manipulated further in \cite{supernum}, where it was cast into a
somewhat more convenient form.  The notations and conventions of
\cite{clps} and \cite{supernum} have both been improved and updated in
their latest hep-th versions.  The reader is referred to these papers
for background material and additional information on notation and
conventions.  The type IIB action as obtained in \cite{supernum} takes
the form
%%%%%
\crampest
\bea
{\cal L} &=& -\ft12 \sqrt{-h}\, h^{ij}\, \del_i X^\mu\, \del_j X^\nu\,
           g_{\mu\nu} + \ft12\ep^{ij}\, \del_i X^\mu\, \del_j X^\nu\,
        B_{\mu\nu} \nn\\
&&+ \im\, \del_i X^\mu\, \btheta\, \gamma^{ij}\, \varrho_0\,
\Gamma_\mu\, D_j\,\theta
-\ft{\im}8 \del_i X^\mu\, \del_j X^\nu\, \btheta\, \gamma^{ij}\,
\varrho_1\, \Gamma_\mu{}^{\rho\sigma}\,
\theta\, G_{\nu\rho\sigma} \label{2baction}\\
&& + \ft{\im}{8} e^{\phi}\, \del_i X^\mu\, \del_j X^\nu \,
\btheta\, \gamma^{ij}\, \Gamma_\mu\,
[ \Gamma^\rho\, \del_\rho\chi + \ft16\varrho_2\,
\Gamma^{\rho_1\rho_2\rho_3}\, F_{\rho_1\rho_2\rho_3}
+\ft1{240} \Gamma^{\rho_1\cdots \rho_5}\,
F_{\rho_1\cdots \rho_5}] \,\Gamma_\nu\,  \theta \,,\nn
\eea
\uncramp
%%%%%
where $G_\3=dB_\2$ is the NS-NS 3-form, $\phi$ and $\chi$ are the
dilaton and axion, and $F_\3$ and $F_\5$ are the R-R 3-form and
self-dual 5-form, and we have defined
%%%%%
\be
\gamma^{ij}\equiv \sqrt{-h}\, h^{ij} - \ep^{ij}\, \varrho_2\,.
\ee
%%%%%
Hence, the conjugate $\btheta$ is defined by $\btheta=\theta^\dagger\,
\Gamma_0\, 
\varrho_0$, and the ``worldsheet'' Dirac matrices are defined by $\varrho_0  
= -\im\, \tau_2$, $\varrho_1=\tau_1$, giving $\varrho_2=\tau_3$, where
the $\tau_a$ are the usual Pauli matrices.
   
   After some algebraic manipulations, we can show that the action 
following from (\ref{2baction}) is invariant under local kappa
transformations defined by
%%%%%
\be
\delta\, \theta = (1+\Gamma)\, \kappa\,,\qquad 
\delta\, X^\mu = \im\, \btheta\, \Gamma^\mu\, \varrho_0\, \delta\, 
\theta\,,\label{2bkappa}
\ee
%%%%%
where the matrix $\Gamma$ is defined by
%%%%%
\be
\Gamma= \fft1{2\sqrt{-h}}\, \ep^{ij}\, \del_i X^\mu\, \del_j X^\nu
\, \Gamma_{\mu\nu}\, \varrho_2\,.\label{Gammaexp}
\ee
%%%%%
(Note that to the quadratic order in fermions to which we are working,
it suffices to replace the usual pulled-back supervielbeins $\Pi_i^m
\equiv \del_i Z^M\, \, E^m_M$ by $\del_i X^\mu\, e_\mu^m$ in the 
expression for $\Gamma$, leading to (\ref{Gammaexp}).)

    We can also show that the action following from (\ref{2baction})
is invariant under rigid spacetime supersymmetry transformations
%%%%%
\be
\delta \, \theta = \ep\,,\qquad \delta\, X^\mu = -\im\, \btheta\, 
\Gamma^\mu\, \varrho_0\, \ep\,,\label{eptrans}
\ee
%%%%%
where $\ep$ is a Killing spinor in the bosonic background spacetime.
Specifically, we can show that under (\ref{eptrans}), the Lagrangian 
(\ref{2baction}) varies to give, up to total derivatives,   
%%%%%
\be
\delta\, {\cal L} = 2\im\, \del_i X^\mu\, \del_j X^\nu\, 
\btheta\, \gamma^{ij}\, \varrho_0\, \Gamma_\mu\,  
{\cal D}_\nu\, \ep\,,
\ee
%%%%%
where
%%%%%
\crampest
\be
{\cal D}_\mu = \nabla_\mu + \ft18 G_{\mu\rho\sigma}\, \varrho_2\, 
\Gamma^{\rho\sigma} - \ft18 e^\phi\, \Big(\varrho_0\, \Gamma^\sigma\, 
\del_\sigma\chi - \ft16 \varrho_1\, \Gamma^{\sigma_1\sigma_2\sigma_3}\, 
F_{\sigma_1\sigma_2\sigma_3}\, + \ft1{240} 
\varrho_0\, \Gamma^{\sigma_1\cdots \sigma_5}\, F_{\sigma_1\cdots 
\sigma_5}\Big)\, \Gamma_\mu\,.\label{cald2b}
\ee
\uncramp
%%%%%
Here $\nabla_\mu\equiv \del_\mu + \ft14\omega_\mu^{mn}\, \Gamma_{mn}$.
Note that ${\cal D}_\mu$ is precisely the supercovariant derivative that
appears in the spacetime supersymmetry transformation of the gravitini
in type IIB supergravity, and so the Green-Schwarz action following from
(\ref{2baction}) is invariant under a rigid supersymmetry (\ref{eptrans})
where $\ep$ is a Killing spinor, satisfying ${\cal D}_\mu\, \ep=0$.

   It is worth remarking that if we define
%%%%%
\crampest
\bea
\wtd{\cal D}_i&=&D_i + \del_i X^\mu\times\\
&& \!\!\!\!\!\! \Big[ \ft18 G_{\mu\rho\sigma}\, 
\varrho_2\, 
\Gamma^{\rho\sigma} - \ft18 e^\phi\, \Big(\varrho_0\, \Gamma^\sigma\, 
\del_\sigma\chi - \ft16 \varrho_1\, \Gamma^{\sigma_1\sigma_2\sigma_3}\, 
\Gamma_{\sigma_1\sigma_2\sigma_3}\, + \ft1{240} 
\varrho_0\, \Gamma^{\sigma_1\cdots \sigma_5}\, F_{\sigma_1\cdots 
\sigma_5}\Big)\, \Gamma_\mu\Big]\,,\nn
\eea
\uncramp
%%%%%
which when acting on a spinor function of the $X^\mu$ coordinates such
as $\ep$ reduces to $\wtd{\cal D}_i\, \ep= \del_i X^\mu\, {\cal
D}\mu\, \ep$, we can write the Lagrangian (\ref{2baction}) as
%%%%%
\be
{\cal L} =  -\ft12 \sqrt{-h}\, h^{ij}\, \del_i X^\mu\, \del_j X^\nu\,
           g_{\mu\nu} + \ft12\ep^{ij}\, \del_i X^\mu\, \del_j X^\nu\,
        B_{\mu\nu}  + \im\, \del_i X^\mu\, \btheta\, \gamma^{ij}\, 
\varrho_0\, \Gamma_\mu\, \wtd{\cal D}_j\, \theta\,.\label{2blagsimp}
\ee
%%%%%
Thus the type IIB string action obtained in \cite{clps} and
\cite{supernum} does indeed satisfy this property, despite the doubts
on this account that were raised in \cite{cohakewa}.

    Note that if we define the doublet of Majorana-Weyl 32-component spinors 
%%%%%
\be
\ep=\pmatrix{\ep_1\cr \ep_2}\,,
\ee
%%%%%
where the upper and lower components correspond to the positive
and negative eigenstates of the worldsheet chirality matrix $\varrho_2$,
and then define the complex Weyl spinor $\varepsilon\equiv \ep_1+\im\, 
\ep_2$, then we have the equivalences
%%%%%
\be
\varrho_2(\ep)\leftrightarrow \varepsilon^*\,,\qquad 
\varrho_0(\ep)\leftrightarrow -\im\, \varepsilon\,,\qquad 
\varrho_1(\ep)\leftrightarrow \im\, \varepsilon^*\,.
\ee
%%%%%
In terms of this complex Weyl notation, the gravitino transformation
rule $\delta\psi_\mu = {\cal D}_\mu\, \ep$ with ${\cal D}_\mu$ given by
(\ref{cald2b}) therefore becomes
%%%%%
\bea
\delta\psi_\mu &=&  \nabla_\mu \varepsilon + 
\ft18 G_{\mu\rho\sigma}\, 
\Gamma^{\rho\sigma}\,\varepsilon^*\nn\\
&&
 +\ft{\im}8 e^\phi\, \Big(\Gamma^\sigma\, \Gamma_\mu\, 
\del_\sigma\chi\, \varepsilon + 
\ft{1}6 \, \Gamma^{\sigma_1\sigma_2\sigma_3}\,\Gamma_\mu\,  
F_{\sigma_1\sigma_2\sigma_3}\, \varepsilon^* + \ft{1}{240} 
\, \Gamma^{\sigma_1\cdots \sigma_5}\, \Gamma_\mu\, F_{\sigma_1\cdots 
\sigma_5}\, \varepsilon\Big)\,,\label{cald2b2}
\eea
%%%%%
where $\psi_\mu$ is now a complex Weyl spinor-vector, defined
analogously to $\varepsilon$.

\subsection{Type IIB worldsheet supersymmetry in the light-cone gauge}
\label{2blightcone}

    The treatment of kappa symmetry and supersymmetry for
the type IIB Green-Schwarz action in light-cone gauge closely 
parallels the type IIA discussion in section \ref{2alightcone}, so we shall
just summarise the key results. The essential change is that the matrix
$\Gamma$ appearing in the kappa transformation rule (\ref{2bkappa})
is now given by
%%%%%
\be
\Gamma= \fft1{2\sqrt{-h}}\, \ep^{ij}\, \Pi_i^m\, \Pi_j^m\, \Gamma_{mn}\, 
\varrho_2\,
\ee
%%%%%
rather than (\ref{Gammaexp}).  Following the same steps as in section 
\ref{2alightcone}, we impose the light-cone condition $\Gamma_-\, \theta=0$,
and so the compensating kappa transformation that maintains this gauge,
analogous to (\ref{lcpres}), is
%%%%%
\be
Q\, \Gamma_-\, (\kappa+\ep) + (\Gamma_-\, \Gamma_+ +\Gamma_-\, P)\, 
\varrho_2\, \kappa=0\,.\label{lcpres2b}
\ee
%%%%%
We then find that the residual transformations of $\Gamma_+\, \theta$,
analogous to the type IIA result (\ref{residualsusy}), are given by
%%%%%
\be
\delta(\Gamma_+\, \theta) = \ft12( P + Q\, \varrho_2)\, \Gamma_+\, 
\Gamma_-\, \ep + \Gamma_+\, \ep\,.\label{residualsusy2b}
\ee
%%%%%

    As in the type IIA case, we can see that the residual
linearly-realised supersymmetries are parametrised by Killing spinors
$\ep$ that are subject to the projection condition
%%%%%
\be
\Gamma_+\, \ep=0\,,\label{linsusycon2b}
\ee
%%%%%
whilst the remaining Killing spinors subject to the opposite projection
condition $\Gamma_-\, \ep=0$ correspond to inhomogeneous shift
transformations.  Thus we see that here for type IIB we have the same
result as in type IIA, namely that the 16 ordinary Killing spinors in
any pp-wave solution, which are subject to the projection $\Gamma_-\,
\ep=0$, are associated with inhomogeneous shift symmetries on the 
string worldsheet.  By contrast, any supernumerary Killing spinors,
which are subject to the projection condition $\Gamma_+\, \ep=0$, are
associated with linearly-realised worldsheet supersymmetries.

\section{Worldsheet Supersymmetry in pp-wave Backgrounds}

\subsection{Killing spinors in pp-wave backgrounds}\label{ppwaveks}

    Before considering pp-waves of the type that arise in Penrose
limits and their generalisations, it is helpful to begin by recalling
the situation for purely gravitational wave solutions in
supergravities, where only the metric itself takes a non-trivial form.
The solutions of this type that we shall consider are given by
%%%%%
\be
ds^2 = 2 dx^+\, dx^- + K\, {dx^+}^2 + dz_i^2\,,\label{genwave}
\ee
%%%%%
where $K$ is a function of the $z_i$.  ($K$ could also be allowed to
depend on $x^+$, but we shall not consider this here.)  The condition
for the Ricci tensor to vanish is that
%%%%%
\be
\square\, K=0\,,\label{keq1}
\ee
%%%%%
where $\square$ is the Laplacian $\del_i^2$ on the flat transverse
space whose coordinates are the $z_i$. 

    If this Ricci-flat solution is taken in the context of a
supergravity theory, the supersymmetry transformation rules will imply
that the background is supersymmetric for spinor parameters $\ep$ that
satisfy $\nabla\, \ep=0$, where $\nabla\equiv d + \ft14 \omega^{AB}\,
\Gamma_{AB}$ is the Lorentz-covariant exterior derivative.  It is
easily seen that in the background (\ref{genwave}), this derivative
is given by
%%%%%
\be
\nabla = d + \ft14 K_i\, \, dx^+\, \Gamma_{-i}\,,
\ee
%%%%%
where $K_i\equiv \del_i\, K$.  A
Killing spinor $\ep$ satisfies $\del_+\,
\ep +\ft14 K_i\, \Gamma_{-i}\, \ep=0$, $\del_-\, \ep=0$ and $\del_i\,
\ep=0$, from which it is straightforward to show that $\ep$ must be
constant, and satisfy the projection condition 
%%%%%
\be
\Gamma_-\, \ep=0\,.\label{gammaminusproj}
\ee
%%%%%

    The conclusion from the above discussion is that whenever we have
a purely gravitational wave solution in supergravity, there will be
Killing spinors whose total number is precisely one half of the number
that would arise in a flat background, on account of the projection 
condition (\ref{gammaminusproj}).  This counting of Killing spinors is
independent of the specific details of the harmonic function $K$.

    In the pp-wave solutions that have been considered recently,
arising as Penrose limits of AdS$\times$Sphere solutions in
supergravity, there are additional non-trivial contributions to the
bosonic background, aside from the gravitational wave metric
(\ref{genwave}) itself.  In fact in all cases, the extra fields
involved in the recently-considered pp-wave solutions take the form of
constant background values for antisymmetric tensor field strengths in the
supergravity theory.  For a $p$-index field strength $F_p$, this background
has the structure
%%%%%
\be
F_p = dx^+ \wedge \Phi_{p-1}\,,\label{fform}
\ee
%%%%%
where $\Phi_{p-1}$ is a constant $(p-1)$-form in the transverse space.
The previous homogeneous equation (\ref{keq1}) for $K$ now becomes
%%%%%
\be
\square\, K = -k\, |\Phi_{p-1}|^2\,,\label{keq2}
\ee
%%%%%
where $k$ is a constant.
 
   In the transformation rules for the gravitini and spin-$\ft12$ fields
in the supergravity theory, these field strengths give contributions
of the form
%%%%%
\bea
\delta\, \psi_\mu &=& \nabla_\mu \, \ep + c_1\, F_{\mu\nu_1\cdots 
 \nu_{p-1}}\, \Gamma^{\nu_1\cdots \nu_{p-1}}\, \ep + 
c_2\, F^{\nu_1\cdots \nu_p}\, \Gamma_{\mu\nu_1\cdots \nu_{p}}\,\ep
+\cdots \,,\nn\\
\delta\, \chi &=& c_3\, F_{\nu_1\, \cdots \nu_p}\, \Gamma^{\nu_1\cdots 
\nu_p}\,\ep + \cdots\,.\label{susyrules}
\eea
%%%%%

    It is clear that the contribution of a field strength
(\ref{fform}) in the supersymmetry transformation rules for $\delta\,
\psi_-$, $\delta\, \psi_i$ and $\delta\, \chi$ will all involve a
$\Gamma_-$ projection acting on $\ep$, and thus these terms will all
vanish if $\ep$ again satisfies the projection condition
(\ref{gammaminusproj}).  There can be a non-vanishing contribution
only in $\delta\, \psi_+$, which means that the previous equation
$\del_+\, \ep=0$ in the purely gravitational case now becomes an
equation determining the $x^+$ dependence of $\ep$.  The upshot is
that all the Killing spinors in the previous purely gravitational wave
background will survive, possibly now with $x^+$ dependence, in the
more general solutions with constant field strengths.  In other
words, in the new pp-wave solutions there will always exist Killing
spinors satisfying the projection condition (\ref{gammaminusproj}), whose
number is again precisely one half of the number of Killing spinors for
a flat background.  In papers \cite{clppp,supernum}, the Killing
spinors that exist for arbitrary choices of $K$ solving the bosonic
equation (\ref{keq2}) were referred to as {\it Standard Killing Spinors}.  

   In the recently-considered pp-wave solutions, there can exist
further Killing spinors over and above the standard ones.  These can
arise only if the constant form $\Phi_{p-1}$ and the solution to
(\ref{keq2}) for $K$ are chosen to have a very special form.  In
particular, $K$ must be taken to be a specific purely quadratic
function of the transverse coordinates $z_i$.  By contrast to the
standard Killing spinors described above, these additional Killing 
spinors, if they occur, are subject to the opposite projection condition,
namely 
%%%%%
\be
\Gamma_+\, \ep=0\,.\label{gammaplusproj}
\ee
%%%%%
These Killing spinors were referred to as {\it Supernumerary Killing 
Spinors} in \cite{clppp,supernum}.  

   To summarise, if we consider pp-waves in a supergravity theory that
has $N$ supercharges, there will always exist $\ft12 N$ standard Killing
spinors, which satisfy the projection condition (\ref{gammaminusproj}).
There may in addition exist a number $N_{\rm sup}$ of 
supernumerary Killing spinors, with $0\le N_{\rm sup}\le \ft12 N$, 
which satisfy the opposite (and orthogonal) projection condition 
(\ref{gammaplusproj}).  The number that occur depends on the details
of the constant form $\Phi_{p-1}$, and the choice of the quadratic function
$K$.   

\subsection{Linearly-realised worldsheet supersymmetries}

    We saw in sections (\ref{2alightcone}) and (\ref{2blightcone}) 
that when one imposes the light-cone gauge conditions
%%%%%
\be
X^+=\tau\,,\qquad \sqrt{-h}\, h^{ij}= \eta^{ij}\, \qquad 
\Gamma_-\, \theta=0
\ee
%%%%%
on the covariant Green-Schwarz type IIA or IIB actions, the local
kappa symmetry and rigid spacetime supersymmetry transmute into a
rigid worldsheet supersymmetry of the light-cone string action.  To be
more precise, we saw that there is an $N=1$ linearly-realised worldsheet
supersymmetry corresponding to every Killing spinor in the
ten-dimensional target spacetime that satisfies the projection 
condition
%%%%%
\be
\Gamma_+\, \ep=0\,.
\ee
%%%%%

   In view of the discussion of Killing spinors in pp-wave backgrounds
given in \cite{clppp,supernum}, which was summarised in section
(\ref{ppwaveks}) above, we see therefore that the 16 standard Killing
spinors in any type IIA or type IIB pp-wave background will never give
rise to linearly-realised worldsheet supersymmetries, since they
satisfy instead the orthogonal projection condition
(\ref{gammaminusproj}).  By contrast, every supernumerary Killing
spinor, since it satisfies the projection condition
(\ref{gammaplusproj}), will give rise to a linearly-realised
worldsheet supersymmetry.

    This result was in fact foreseen in \cite{clppp,supernum}.  The
argument used there was based on the observation that in the
light-cone string action the masses of the bosonic fields $X^I$ and
the fermionic fields $\theta$ are in general unrelated, in a pp-wave
background.  It is manifest that if the boson and fermion masses are
unequal then there cannot be any linearly-realised worldsheet
supersymmetry.  Only in the case of special pp-wave backgrounds where
there exist supernumerary Killing spinors does one find that there is
a precise matching of boson and fermion masses.  This provided
circumstantial evidence in \cite{clppp,supernum} for the connection
between supernumerary Killing spinors and linearly-realised worldsheet
supersymmetries, which we have now made precise in this paper.

\section{Green-Schwarz Actions for pp-waves in Physical Gauge}
\label{gsphysgauge}

\subsection{Type IIA worldsheet supersymmetries in the physical gauge}
\label{2aphysgauge}

   Let us contrast the light-cone analysis given above with what happens
if we instead choose a physical gauge for the type IIA Green-Schwarz action.
This gauge can be taken to be
%%%%%
\be
X^0=\tau\,,\qquad X^9=\sigma\,,\qquad (1+\Gamma_*)\, \theta=0\,,
\ee
%%%%%
where the matrix $\Gamma_*$ is defined by
%%%%%
\be
\Gamma_*\equiv \Gamma_1\cdots \Gamma_8\,,\label{gammastar}
\ee
%%%%%
\ie it is the product of the eight transverse-space Dirac matrices.
This fermionic gauge condition is motivated by considering the form of
the matrix $\Gamma$ in the field-independent static limit in the
bosonic part of the physical gauge, where all the coordinates $X^\mu$
except for those set equal to the worldsheet coordinates vanish. From
(\ref{2agamma}) we see that $\Gamma$ then becomes just $\Gamma_{09}\,
\Gamma_{11}$, which is precisely $\Gamma_*$.  Thus the leading-order
form of the kappa symmetry transformation is $\delta\, \theta =
(1+\Gamma_*)\, \kappa$, and so we can expect to be able to use this
symmetry to set\footnote{One might instead use the kappa
transformation to set $(1+\Gamma)\, \theta=0$ as a gauge condition,
but this would be less convenient since it is field dependent.}
$(1+\Gamma_*)\, \theta =0$.  In fact we shall show in section
\ref{2astringwave} that this physical-gauge fixing condition is
precisely what arises naturally in a string/pp-wave intersecting
supergravity solution.

   We now find it helpful to decompose the Dirac $32\times 32$
matrices as\footnote{Each entry is an $8\times 8$ matrix here.}
%%%%%
\be
\Gamma_0= \pmatrix{0&1&0&0\cr -1&0&0&0\cr 0&0&0&1\cr 0&0&-1&0}\,,\quad
\Gamma_9= \pmatrix{0&1&0&0\cr 1&0&0&0\cr 0&0&0&1\cr 0&0&1&0}\,,\quad
\Gamma_I = \pmatrix{0&0&\beta_I &0\cr 0&0&0&-\beta_I\cr
           \td\beta_I &0&0&0\cr 0&-\td\beta_I&0&0}\,,\label{2adecomp}
\ee
%%%%%
where $\beta_I=(\beta_{I'},\beta_8)=(\td\gamma_{I'},\im)$
and $\td\beta_I=(\td\beta_{I'},\td\beta_8)=(\td\gamma_{I'},-\im)$
are the Van der Waerden symbols of eight dimensions, with $\td\gamma_{I'}$ 
being the $8\times8$ Dirac matrices in seven dimensions.  With these
conventions, we have
%%%%%
\be
\Gamma_* = \pmatrix{1&0&0&0\cr 0&1&0&0\cr 0&0&-1&0\cr 0&0&0&-1}\,,\quad
\Gamma_{11}=\pmatrix{1&0&0&0\cr 0&-1&0&0\cr 0&0&-1&0\cr 0&0&0&1}\,.
\label{gamma*11}
\ee
%%%%%

   The matrix $\Gamma$ that appears in the kappa transformations can 
be written as 
%%%%%
\be
\Gamma= \ft12 M^{mn}\, \Gamma_{mn}\, \Gamma_{11}=-\oneone + A + B\,,
\ee
%%%%%
where
%%%%%
\bea
A &\equiv & \oneone + M^{09}\, \Gamma_{09}\, \Gamma_{11} + \ft12
    M^{IJ}\, \Gamma_{IJ}\, \Gamma_{11}\,,\nn\\
B&\equiv & = M^{0I}\, \Gamma_{0I}\, \Gamma_{11} 
   + M^{I9}\, \Gamma_{I9}\, \Gamma_{11}\,.
\eea
%%%%%
This splitting of the terms in $\Gamma$ is chosen so that
%%%%%
\be
[A,\Gamma_*]=0\,,\qquad \{B,\Gamma_*\}=0\,.\label{ABcom}
\ee
%%%%%
The quantities $M^{mn}$ are given by
%%%%%
\bea
M^{09} &=& \fft1{\sqrt{-h}}\, (\Pi_0^0\, \Pi_1^9 -\Pi_1^0\, \Pi_0^9)\,,\nn\\
M^{0I} &=& \fft1{\sqrt{-h}}\, (\Pi_0^0\, \Pi_1^I -\Pi_1^0\, \Pi_0^I)\,,\nn\\
M^{I9} &=& \fft1{\sqrt{-h}}\, (\Pi_0^I\, \Pi_1^9 -\Pi_1^I\, \Pi_0^9)\,,\nn\\
M^{IJ} &=& \fft1{\sqrt{-h}}\, (\Pi_0^I\, \Pi_1^J -\Pi_1^I\, \Pi_0^J)\,.
\eea
%%%%%

   We now proceed with the physical gauge fixing.  Requiring that
$(1+\Gamma_*)\, \theta=0$ be preserved under combined supersymmetry and 
kappa transformations implies
%%%%%
\be
A\, (1+ \Gamma_*)\, \kappa + B\, (1-\Gamma_*)\, \kappa + 
(1+\Gamma_*)\, \ep=0\,,\label{pluscon}
\ee
%%%%%
and then the transformation of $(1-\Gamma_*)\, \theta$ is given by
%%%%%
\be
(1-\Gamma_*)\, \delta\, \theta= 
A\, (1- \Gamma_*)\, \kappa + B\, (1+\Gamma_*)\, \kappa + 
(1-\Gamma_*)\, \ep\,.\label{minuscomb}
\ee
%%%%%
The commutation properties of $A$ and $B$ given in (\ref{ABcom}) imply
that we can write them in $16\times 16$ block-matrix form, for which
$\Gamma_*$ in (\ref{gamma*11}) becomes
%%%%
\be
\Gamma_*=\pmatrix{\oneone_{16} &0\cr 0 & -\oneone_{16}}\,,\label{16gamma*}
\ee
%%%%%
 as 
%%%%%
\be
A=\pmatrix{a &0\cr 0&\td a}\,,\qquad B= \pmatrix{0& b\cr \td b &0}\,.
\label{abblock}
\ee
%%%%%
Defining 
%%%%%
\be
\theta=\pmatrix{\theta_1\cr\theta_2}\,,\qquad
\kappa=\pmatrix{\kappa_1\cr\kappa_2}\,,\qquad
\ep=\pmatrix{\ep_1\cr\ep_2}\,,
\ee
%%%%%
where the upper and lower 16 components are the projections under
$\ft12(1\pm\Gamma_*)$ with $\Gamma_*$ given in (\ref{16gamma*}),
we therefore find that (\ref{pluscon}) and (\ref{minuscomb}) become
%%%%
\bea
0&=& a\, \kappa_1 + b\, \kappa_2 + \ep_1\,,\nn\\
\delta\, \theta_2 &=& \td a\, \kappa_2 + \td b\, \kappa_0 + \ep_2\,.
\label{abvars}
\eea
%%%%%

   Note that the physical-gauge choice means that $\Pi_0^0$ and $\Pi_1^9$
are of order 1, whilst all the other $\Pi_i^m$ are of first order in
physical fields, plus higher terms.  This means that up to linear order
in coordinates, expanding around the static background $X^0=\tau,
X^9=\sigma, X^I=0, \theta=0$, we have
%%%%%
\be
M_{09}=1\,,\quad M^{0I}= \del_1 X^I\,,\quad M^{I9}= \del_0 X^I\,,\quad
M^{IJ} =0\,.
\ee
%%%%%
In particular, we note that the matrix $A$ has the form
%%%%
\be
A=\pmatrix{2&0&0&0\cr 0&2&0&0\cr 0&0&0&0\cr 0&0&0&0} + \hbox{linear
terms}\,,
\ee
%%%%
which means that its upper $16\times 16$ block denoted by $a$ in
(\ref{abblock}) is invertible.  Thus we can solve the first equation 
in (\ref{abvars}) for $\kappa_1$, and substitute into the second.  This
gives
%%%%%
\be
\delta\, \theta_2 = -\td b\, a^{-1}\, \ep_1 + \ep_2 + (\td a
- \td b\, a^{-1}\, b)\, \kappa_2\,.\label{nextvar}
\ee
%%%%%

   Next, we note from the fact that $\Gamma^2=1$ that we shall have
$(A+B)(A+B-2)=0$, which from (\ref{abblock}) gives
%%%%%
\be
a(a-2)+ b\td b=0\,,\quad ab + b(\td a-2)=0\,, \quad
\td b (a-2) + \td a\td b=0\,,
\quad \td b b + \td a(\td a-2)=0\,.
\ee
%%%%%
Multiplying the second equation by $\td b a^{-1}$ from the left, and 
subtracting the fourth, whilst noting that $\td a$ is of first order in 
fields and that therefore $(\td a-2)$ is invertible, we deduce
that $\td a = \td b\, a^{-1}\, b$.  This shows that $\kappa_2$ drops
out completely from (\re{nextvar}), and hence that we have
%%%%%
\be
\delta\theta_2 = - \td b\, a^{-1}\, \ep_1 + \ep_2\,.\label{var2}
\ee
%%%%% 

   Finally, we can see from the previous definitions that up to linear 
order in fields (for which, since $\td b$ is of linear order we
can take $a^{-1}$ in (\ref{var2}) to be $\ft12$), 
%%%%%
\be
\delta\, \theta_2 = -\ft12\td b\, \ep_1 + \ep_2 \,,
\ee
%%%%%
where
%%%%%
\be
\td b= \pmatrix{ 0 & \td\beta_I\, \del_- X^I\cr
               -\td\beta_I\, \del_+ X^I &0}\,.
\ee
%%%%%
Here, we have defined the worldsheet derivatives 
$\del_\pm = \del_1 \pm \del_0$.  We see, therefore, that $\ep_1$
parametrises linearly-realised supersymmetries, while $\ep_2$
describes an inhomogeneous shift symmetry.  In summary, therefore, we 
have shown that in the physical gauge the parameter $\ep$ of the
linearly-realised worldsheet supersymmetry is subject to the projection
condition
%%%%%
\be
(1-\Gamma_*)\, \ep=0\,.\label{starcon}
\ee
%%%%%

    The projection condition (\ref{starcon}) for linearly-realised
supersymmetries in the physical gauge is very different from the
analogous projection condition (\ref{linsusycon}) that we obtained in
the light-cone gauge.  In particular, we find that whereas the 16
standard Killing spinors in any pp-wave background are all
incompatible with the light-cone projection (\ref{linsusycon}), in the
physical gauge some of the standard Killing spinors are compatible
with the projection (\ref{starcon}).  In fact, as we shall see in
detail in section \ref{sugrasec}, there are both standard and
supernumerary Killing spinors in pp-waves that are compatible with the
projection (\ref{starcon}) in the physical gauge.  This means that in
the physical gauge, there can be linearly-realised worldsheet
supersymmetries even in a pp-wave background that has only the 16
standard Killing spinors.

\subsection{Type IIB worldsheet supersymmetries in the physical gauge}
\label{2bphysgauge}

   The imposition of a physical gauge condition in the type IIB string
proceeds very analogously to the discussion for type IIA 
in section \ref{2aphysgauge}.  Again, one can determine a suitable
choice of fermionic gauge condition by starting with a static string
configuration where $X^0=\tau$, $X^9=\sigma$ and all other bosonic
coordinates vanish.  The full expression 
%%%%%
\be
\Gamma= \fft1{2\sqrt{-h}}\, \ep^{ij}\, \Pi_i^m\, \Pi_j^n
\, \Gamma_{mn}\, \varrho_2\,.\label{2bGammaexp}
\ee
%%%%%
then reduces to 
%%%%%
\be
\Gamma\longrightarrow \Gamma_\# \equiv \Gamma_{09}\, \varrho_2\,,
\ee
%%%%%
and so we are led to impose
%%%%%
\be
X^0=\tau\,,\qquad X^9\equiv \sigma\,,\qquad (1+\Gamma_\#)\, \theta=0
\ee
%%%%%
as the physical gauge condition.  We shall show later that this
fermionic projection agrees precisely with the one that is encountered
from the string component in a type IIB string/pp-wave intersection.

   As in the type IIA discussion in section \ref{2aphysgauge}, it is
convenient to make a choice of Dirac matrix basis that is adapted to
the gauge condition.  Specifically, we choose a basis where
$\Gamma_\#$ is diagonal, implying that its non-vanishing components
will then be $\pm1$, and with the positive components appearing in the
upper left-hand part of the matrix.  We can start from the basis 
(\ref{2adecomp}), and first perform a similarity transformation in which 
$\Gamma_{11}$ becomes
%%%%%
\be
\Gamma_{11} = \pmatrix{1&0&0&0\cr 0&1&0&0\cr 0&0&-1&0\cr 0&0&0&-1}\,.
\ee
%%%%%
This can be achieved by exchanging the roles of the second and fourth
rows and columns, leading to
%%%%%
\be
\Gamma_0=\pmatrix{0&0&0&1\cr 0&0&-1&0\cr 0&1&0&0\cr -1&0&0&0}\,,\quad
\Gamma_9=\pmatrix{0&0&0&1\cr 0&0&1&0\cr 0&1&0&0\cr 1&0&0&0}\,,\quad
\Gamma_I=\pmatrix{0&0&\beta_I &0\cr 0&0&0& -\td\beta_I\cr
                  \td\beta_I &0&0&0\cr 0&-\beta_I &0&0}\,.
\ee
%%%%%
Note that the Dirac matrix combinations $\Gamma_{mn}$ that appear in 
$\Gamma$ all have components only in the upper left and lower right
$16\times 16$ blocks.  

   The spinors $\theta$ in type IIB are chiral, and so only the upper
16 components in the new basis will be non-vanishing.  We can thus
focus attention on these components, and now consider the resulting 
$32\times 32$ matrices in the tensor product with the worldsheet Dirac
matrices $\varrho_a$.  In particular, with $\varrho_2=\tau_3$ we shall 
have
%%%%%
\bea
&&\Gamma_\# = \Gamma_{09}\, \varrho_2 \longrightarrow 
\pmatrix{1&0&0&0\cr 0&-1&0&0\cr 0&0&-1&0\cr 0&0&0&1}\,,\quad
\Gamma_{0I}\, \varrho_2 \longrightarrow 
\pmatrix{0&-\beta_I &0 &0\cr -\td\beta_I &0&0&0\cr
         0&0&0&\beta_I\cr 0&0&\td\beta_I&0}\,,\\
&&\Gamma_{I9}\, \varrho_2 \longrightarrow 
\pmatrix{0&\beta_I &0 &0\cr -\td\beta_I &0&0&0\cr
         0&0&0&-\beta_I\cr 0&0&\td\beta_I&0}\,,\
\Gamma_{IJ}\, \varrho_2\longrightarrow 
\pmatrix{\beta_I\, \td\beta_J &0&0&0\cr 0&\td\beta_I\, \beta_J &0&0\cr
0&0&-\beta_I\, \td\beta_J &0\cr 0&0&0&-\td\beta_I\, \beta_J}\,.\nn
\eea
%%%%%
 
   Finally, we apply a similarity transformation whose effect is to
exchange the roles of the second and fourth rows and columns in these
matrices, leading to
%%%%%
\bea
&&\Gamma_\# = \Gamma_{09}\, \varrho_2 \longrightarrow 
\pmatrix{1&0&0&0\cr 0&1&0&0\cr 0&0&-1&0\cr 0&0&0&-1}\,,\quad
\Gamma_{0I}\, \varrho_2 \longrightarrow 
\pmatrix{0&0 &0 &-\beta_I\cr 0&0&\td\beta_I&0\cr
         0&\beta_I&0&0\cr -\td\beta_I&0&0&0}\,,\\
&&\Gamma_{I9}\, \varrho_2 \longrightarrow 
\pmatrix{0&0 &0 &\beta_I\cr 0&0&\td\beta_I&0\cr
         0&-\beta_I&0&0\cr -\td\beta_I&0&0&0}\,,\ 
\Gamma_{IJ}\, \varrho_2\longrightarrow 
\pmatrix{\beta_I\, \td\beta_J &0&0&0\cr 0&-\td\beta_I\, \beta_J &0&0\cr
0&0&-\beta_I\, \td\beta_J &0\cr 0&0&0&\td\beta_I\, \beta_J}\,.\nn
\eea
%%%%%
In this basis, when we construct $\Gamma\equiv -\oneone + A+B$ with
%%%%%
\bea
A&\equiv& \oneone + M^{09}\, \Gamma_{09}\, \varrho_2 +\ft12 M^{IJ}\, 
\Gamma_{IJ}\, \varrho_2\,,\nn\\
B&\equiv& M^{0I}\, \Gamma_{0I}\, \varrho_2 + M^{I9}\, \Gamma_{I9}\, 
\varrho_2\,,
\eea
%%%%%
the fact that $[A,\Gamma_\#]=0$ and $\{B,\Gamma_\#\}=0$ implies
that in $16\times 16$ block form we shall have
%%%%%
\be
A=\pmatrix{a &0\cr 0&\td a}\,,\qquad
B=\pmatrix{0 &b\cr \td b&0}\,.
\ee
%%%%%
This is closely parallel to the situation for the type IIA string,
although here the $16\times 16$ matrices $(a,\td a, b, \td b)$ are
different from those appearing in (\ref{abblock}). 

   From this point on, the imposition of the physical gauge conditions
proceeds exactly in parallel with the type IIA discussion in section
\ref{2aphysgauge}.  Thus we impose $(1+\Gamma_\#)\, \theta=0$, solve
for the compensating kappa transformation that preserves this gauge,
and thereby arrive at the residual transformations
%%%%%
\be
\delta\,\theta_2 = -\td b\, a^{-1}\, \ep_1 + \ep_2\,,
\ee
%%%%%
where we have written
%%%%
\be
\theta=\pmatrix{\theta_1\cr \theta_2}\,,\qquad \ep=\pmatrix{\ep_1\cr
\ep_2}\,.
\ee
%%%%%
To first order in fields around the static string configuration,
we shall therefore have
%%%%%
\be
\delta\, \theta_2 = -\ft12\td b\, \ep_1+\ep_2\,,
\ee
%%%%%
with
%%%%%
\be
\td b= \pmatrix{0& \beta_I\, \del_- X^I\cr -\td\beta_I\, \del_+ X^I &0}\,.
\ee
%%%%%
The linearly-realised supersymmetries are parametrised by $\ep_1$,
\ie by Killing spinors $\ep$ that are subject to the projection condition
%%%%%
\be
(1-\Gamma_\#)\, \ep=0\,.\label{hashcon}
\ee
%%%%%

   As in the case of type IIA strings, so here for type IIB strings
we see that the projection condition (\ref{hashcon}) for
linearly-realised worldsheet supersymmetries in the physical gauge is
very different from the analogous projection (\ref{linsusycon2b}) in
the light-cone gauge.  Again, this implies that both standard as well
as supernumerary Killing spinors in a pp-wave solution can give rise
to linearly-realised worldsheet supersymmetries in the physical gauge.
This is discussed in more detail in section \ref{sugrasec}.

\subsection{Absence of mass-terms in the physical gauge}

   The forms of the complete Green-Schwarz actions, after imposing the 
physical gauge conditions, are quite complicated and we shall not
present them in detail here.  Instead, we shall focus just on the
sectors where mass terms for the bosons and fermions might arise, in
order to demonstrate that they are in fact absent.

   First, we consider the bosonic sector, and consider the term
%%%%%
\be
{\cal L}_0 = -\ft12 \sqrt{-h}\, h^{ij}\, \del_i X^\mu\, \del_j
X^\nu\,,\label{l0term}
\ee
%%%%%
which is in fact common to both the type IIA and the type IIB
Green-Schwarz actions.  Looking at this sector will be sufficient to
demonstrate the absence of mass terms for the transverse bosonic coordinates
$X^I$ in a pp-wave background.

   In this bosonic sector, we have $h_{ij}= \del_i X^\mu\, \del_j
X^\nu\, g_{\mu\nu}$, and the metric in the pp-wave background is given
in (\ref{genwave}).  The Lagrangian (\ref{l0term}), which can be 
written simply as ${\cal L}_0 =-\sqrt{-h}$, is therefore given by
%%%%%
\bea
{\cal L}_0 &=& -\sqrt{1+2\del_+ X^I\, \del_- X^I - K\, \del_- X^I\, 
\del_+ X^I + (\del_+ X^I\, \del_- X^I)^2 - (\del_- X^I)^2\, 
(\del_+ X^J)^2}\,,\nn\\
&=& -1 -\del_+ X^I\, \del_- X^I +\hbox{interaction terms}\,.
\eea
%%%%%
In particular, we see that, unlike in the light-cone gauge, there are
no mass terms for the $X^I$ transverse coordinates.

   For the fermions, we note that in both the type IIA and type IIB
Green-Schwarz actions, the source of the fermion masses in the
light-cone analysis is the relevant R-R coupling term, which has the form
%%%%%
\be
{\cal L}_{\rm {RR}} = c\, e^{\phi}\, \del_i X^\mu\, \del_j X^\nu
\, \btheta \alpha^{ij}\, \Gamma_\mu\, \Gamma^{\rho_1\cdots \rho_p}\, 
F_{\rho_1\cdots \rho_p}\, \Gamma_\nu\, \theta\,,\label{lrr}
\ee
%%%%%
where $\a^{ij}$ represents either $\beta^{ij}$ in the type IIA case,
given in (\ref{type2alag}) (possibly with an additional $\Gamma_{11}$
factor), or $\gamma^{ij}$ in the type IIB case, given in
(\ref{2baction}) (possibly with an extra factor of $\varrho_2$).
In all cases, the relevant R-R field in the pp-wave solution has the
form $F=\mu\, dx^+\wedge \Phi$.  

   In the physical gauge, purely quadratic fermion terms will
arise when $\del_i X^\mu$ and $\del_j X^\nu$ have their constant
background contributions coming from $\del_+ X^+=1$ and $\del_- X^-=1$, and
so we see that the matrix between the $\btheta$ and $\theta$ fields in
(\ref{lrr}) will involve 
%%%%%
\be
\Gamma_+ \, (\Gamma_-\, W)\, \Gamma_-\qquad \hbox{or}\qquad 
\Gamma_- \, (\Gamma_-\, W)\, \Gamma_+ \,,
\ee
%%%%%
where $W=\fft1{p!}\, \Gamma^{i_1\cdots i_p}\, \Phi_{i_1\cdots i_p}$.
However, both these terms vanish, because $\Gamma_-\,\Gamma_-=0$, and
so we see that in the physical gauge, there are no fermion mass terms.

   We showed in sections \ref{2aphysgauge} and \ref{2bphysgauge} that
there will always be linearly-realised worldsheet supersymmetries for
pp-wave backgrounds in the physical gauge, even when there are only
the 16 ``standard'' Killing spinors and no supernumerary Killing
spinors.  It is satisfactory, therefore, that we have found that there
are no mass terms for either the bosons or the fermions in the
physical gauge, since this is compatible with the linearly-realised
supersymmetry.

\section{Strings in pp-wave Backgrounds: Supergravity Solutions}
\label{sugrasec}

   In this section, we look for solutions in type IIA and type IIB 
supergravity corresponding to fundamental strings in the background of
pp-waves.  These solutions can be viewed as classical supergravity 
realisations of a string action with a pp-wave target spacetime
background.  As discussed in section \ref{gsphysgauge}, the
supergravity solutions describing strings in pp-wave backgrounds will
correspond to string actions that are expressed in the physical gauge.

   The ten-dimensional metrics describing solutions for strings in
pp-wave backgrounds have the same general structure as one finds for
intersecting $p$-brane solutions.  Thus the metric is given by
%%%%%
\be
ds_{10}^2 = H^{-3/4}\, (2 dx^+\, dx^- + K\, {dx^+}^2) + H^{1/4}\, 
    dz_i^2\,,\label{hkmetric}
\ee
%%%%%
where $H$ and $K$ are taken to depend only on the eight transverse 
coordinates $z^i$.  In a standard string/wave intersection, which
could be either in the type IIA or type IIB theory, $H$ and
$K$ would both be harmonic functions and the string source would be
provided by the NS-NS 3-form field strength. The 3-form and dilaton
would be given by
%%%%%
\be
F_\3= dH^{-1}\wedge dx^+\wedge 
          dx^-\,,\qquad \phi = -\ft12 \log H\,.
\ee
%%%%%
For our purposes, we now need to introduce additional form-field
fluxes, just as in the pure pp-wave solutions.  It is not {\it a priori}
obvious that we can still find ``intersecting'' solutions, but in
fact, as we shall show below, we are able to find examples both in
type IIA and type IIB supergravity.

   Before moving to the specific discussions for the type IIA and type
IIB supergravities, it is useful to collect some general results that
are common to the two cases.  

   We introduce the following vielbein 1-forms for the ten-dimensional metric
(\ref{hkmetric}):
%%%%%
\be
e^+ = H^{-3/8}\, dx^+\,,\qquad e^-= H^{-3/8}\, (dx^- + \ft12 K\, dx^+)
\,,\qquad e^i = H^{1/8}\, dz^i\,,
\ee
%%%%%
where the local Lorentz metric in the tangent frame is taken to be
%%%%%
\be
\eta_{+-}=\eta_{-+}= 1\,,\qquad \eta_{ij}=\delta_{ij}\,.
\ee
%%%%%
We find that the torsion-free spin connection is then given by
%%%%%
\bea
&&\omega^+{}_i = -\ft38 H^{-9/8}\, H_i\, e^+\,,\quad
\omega^-{}_i = -\ft38 H^{-9/8}\, H_i \, e^- + \ft12 H^{-1/8}\, K_i\, e^+
\,,\nn\\
&&\omega^i{}_j = -\ft18 H^{-9/8}\, (H_i\, e^j- H_j\, e^i)\,,
\label{2bspincon}
\eea
%%%%%
where $H_i\equiv \del_i\, H$ and $K_i\equiv \del_i\, K$.  It follows
from (\ref{2bspincon}) that the Lorentz-covariant exterior derivative is
$\nabla=d +\ft14\omega^{AB}\, \Gamma_{AB}$, from which one can read off
the vielbein components $\nabla_A$ via $\nabla=e^A\, \nabla_A$, is
given by
%%%%%
\be
\nabla = d -\ft3{16} H^{-9/8}\, H_i\, \Gamma_{+i}\, e^+ - 
\ft3{16} H^{-9/8}\, H_i\, \Gamma_{-i}\, e^- + \ft14 H^{-1/8}\, K_i\, 
\Gamma_{-i}\, e^+ - \ft1{16} H^{-9/8}\, H_i\, \Gamma_{ij}\, e^j\,.
\ee
%%%%% 

   From (\ref{2bspincon}), we find that the frame components of the
Ricci tensor are given by
%%%%%
\bea
&& R_{+-}= +\ft3{8}\,H^{-5/4}\, \square \, H - \ft3{8} H^{-9/4}\,
       H_i\, H_i\,,\quad R_{++}= -\ft12 H^{-1/4}\, \square \, K\,,\quad 
R_{--}=0\,,\nn\\
&& R_{ij} = -\ft18 H^{-5/4}\, \square \, H\, \delta_{ij} + \ft18 H^{-9/4}\, 
H_k\, H_k\, \delta_{ij} - \ft38 H^{-9/4}\, H_i\, H_j\,,\label{riccicomp}
\eea
%%%%%
where $\square\equiv \del_i\del_i$ is the Laplacian in the flat
eight-dimensional transverse space.

\subsection{The type IIB supergravity solution}\label{2bsolsec}

   We seek here a solution describing the intersection of a string and
a pp-wave in the type IIB theory.  There is a standard well-known such
solution for source-free strings and waves, and so we can take this as
our starting point, and look for a generalisation in which the 5-form
is taken to have a constant background value, as in the pp-wave.  Thus
we consider a configuration where the metric is given by
(\ref{hkmetric}), and the field strengths are
%%%%%
\be
F_\5 = \mu\, dx^+ \wedge \Phi_\4\,,\qquad G_\3= dH^{-1}\wedge dx^+\wedge 
          dx^-\,,\qquad \phi = -\ft12 \log H\,,\label{2bsol}
\ee
%%%%%
where $G_\3$ is the NS-NS 3-form, and $\Phi_\4$ is a constant
self-dual tensor in the transverse space $\R^8$, whose coordinates are the
$z^i$.  The novel feature, for an intersection, is the inclusion of
the 5-form term.

   The type IIB equations of motion for the fields that we are taking to
be non-zero are
%%%%%
\bea
&& d(e^{-\phi}\, {*G_\3}) = 0\,,\qquad {*F_\5}=F_\5\,,\qquad 
 dF_\5=0\,,\nn\\
&&R_{AB} = \ft12 \del_A\phi\, \del_B \phi + \ft14 e^{-\phi} 
(G_{ACD}\, G_B{}^{CD} - \!\ft1{12} G_{CDE}\, G^{CDE}\, g_{AB}) +
\ft1{96}\, F_{ACDEF} F_B{}^{CDEF}\,,\nn\\
&&\square\, \phi = -\ft1{12}\, e^{-\phi}\, G_{ABC}\, G^{ABC}\,.
\eea
%%%%%   
After some straightforward algebra, we find that these are satisfied provided
that the functions $H$ and $K$ satisfy
%%%%%
\be
\square \, H = 0\,,\qquad \square \, K = -\ft1{48}\, \mu^2\, |\Phi_\4|^2\,.
\ee
%%%%%
  
   If we set the constant $\mu=0$, we just get back the standard solution 
for the intersection of a string and a wave.  The new feature here is
that we still  obtain a solution if we take $\mu$ to be non-zero.  Thus,
in particular, we have a solution with
%%%%%
\be
K= c_0 -\sum_i \mu_i^2\, z_i^2\,, \qquad H= 1 + \fft{Q}{r^6}\,,
\ee
%%%%%
where $\sum_i \mu_i^2 = \ft1{96}\, \mu^2\, |\Phi_\4|^2$, and $r^2=z_i^2$.
(We could, of course, choose more complicated solutions for $H$ and for
$K$.)

\subsection{Supersymmetry of the type IIB string on a pp-wave}

   The supersymmetry transformation rules for the type IIB gravitini
$\psi_M$ and dilatini $\lambda$, in the background involving the
dilaton, NS-NS 3-form and the 5-form are
%%%%%
\bea
\delta \psi_M &=&\nabla_M \, \ep + \ft{\im}{192} F_{MN_1\cdots N_4}\, 
\Gamma^{N_1\cdots N_4} \, \ep - \ft1{96}\, e^{-\fft12\phi}\, 
G_{NPQ}\, (\Gamma_M{}^{NPQ} - 9 \delta_M^N\, \Gamma^{PQ})\, \ep^*\,,\nn\\
\delta \lambda &=& \Gamma^M\, \del_M\phi\, \ep^* 
-\ft1{12} e^{-\fft12\phi}\, G_{MNP}\, \Gamma^{MNP}\, \ep
\,.\label{2bsusy1}
\eea
%%%%% 
It is helpful to define
%%%%%
\be
U\equiv \ft1{5!} F_{N_1\cdots N_5}\, \Gamma^{N_1\cdots N_5}\,, 
\qquad V\equiv \ft1{3!}\, G_{MNP}\, \Gamma^{MNP}\,,
\ee
%%%%%
in terms of which the gravitino transformation rule becomes
%%%%%
\be
 \delta\psi_M = \nabla_M\, \ep + \ft{\im}{16}\, \{\Gamma_M, U\}\, \ep 
-\ft1{16} e^{-\fft12\phi}\, (\Gamma_M\, V + 2V\, \Gamma_M)\, \ep^*=0\,.
\label{2bsusy2}
\ee
%%%%%
In our background, where $F_\5$ is given in (\ref{2bsol}), we have
%%%%%
\be
U= \mu\, \Gamma_-\, W\,,\qquad W\equiv \ft1{4!}\, \Phi_{ijk\ell}\, 
\Gamma^{ijk\ell}\,.
\ee
%%%%%

    From the dilatino transformation rule in (\ref{2bsusy1}) we obtain
%%%%%
\be
\delta\lambda = -\ft12 H^{-9/8}\, H_i\, \Gamma_i\, (\ep^* + \Gamma_{+-}\, 
\ep)=0\,,
\ee
%%%%%
implying that the string imposes the projection condition
%%%%%
\be
 \ep^* + \Gamma_{+-}\, \ep=0\,.
\ee
%%%%
Using this, we then find that the transverse components $\delta\psi_i=0$
of the gravitino transformation rules imply
%%%%%
\be
\ep=H^{-\fft3{16}}\, \ep_0\,,\qquad \del_i \, \ep_0 - \ft{\im}{16}\, \mu\, 
\Gamma_-\, [\Gamma_i,W]\, \ep_0=0\,.
\ee
%%%%%
From $\delta\psi_+ =0$ we get 
%%%%%
\be
\Gamma_+\, (\ep-\ep^*)=0 \,,\qquad H^{1/2}\, 
\del_+\, \ep  +\ft14 K_i\, \Gamma_-\, \Gamma_i
\, \ep + \ft{\im}8 \mu\, W\, \ep=0\,,
\ee
%%%%%
and finally from $\delta\, \psi_-=0$ we obtain
%%%%%
\be
\Gamma_-\, (\ep+\ep^*)=0 \,,\qquad \del_-\, \ep=0\,.
\ee
%%%%%
  
   These conditions can be summarised as follows.  We have the usual
$\ep=H^{-3/16}\, \ep_0$ factor and the projection conditions
%%%%%
\be
\Gamma_+\, (\ep-\ep^*)=0\,,\qquad \Gamma_-\, (\ep+\ep^*)=0
\ee
%%%%%
for a string in type IIB theory.  The remaining equations
%%%%%
\be
\del_i \, \ep_0 - \ft{\im}{16}\, \mu\, 
\Gamma_-\, [\Gamma_i,W]\, \ep_0=0\,,\quad 
H^{1/2}\, 
\del_+\, \ep_0  +\ft14  K_i\, \Gamma_-\, \Gamma_i
\, \ep_0 + \ft{\im}8 \mu\, W\, \ep_0=0\,,\quad \del_-\, \ep_0=0
\label{2bwavecon}
\ee
%%%%%
are like those for a pp-wave, except for the factor $H^{1/2}$ that
multiplies $\del_+$.  This means that we can only have Killing spinors
that are {\it independent of $x^+$}.  

   Analysis of the content of equations (\ref{2bwavecon}) is
analogous to the discussion for pure pp-waves given in \cite{clppp}.  The
first equation implies that $\ep_0$ can be written as
%%%%%
\be
\ep_0= \Big( 1 + \ft{\im}{16} \, \mu\, [\Gamma_i,W]\, \Gamma_-\Big)\, 
\chi\,,
\ee
%%%%%
where $\chi$ is independent of $z_i$.  In the present context, with a
superimposed string, $\chi$ is therefore a constant spinor.  It then 
follows from (\ref{2bwavecon}) that $\chi$ must satisfy
%%%%%
\be
W\, \chi=0\,,\qquad 
(\mu^2\, z_i\, W^2 + 32 K_i)\, \Gamma_i\, \Gamma_-\, \chi=0\,.
\ee
%%%%%

\subsection{The type IIA supergravity solution}\label{2aswsol}

    We shall look for an intersecting solution analogous to the 
solution (\ref{2bsol}) that we found in the type IIB theory.  We again
take the metric to be (\ref{hkmetric}), and for the field strengths 
we take
%%%%%
\be
F_\4 = \mu\, dx^+ \wedge \Phi_\3\,,\qquad F_\3= dH^{-1}\wedge dx^+\wedge 
          dx^-\,,\qquad \phi = -\ft12 \log H\,,\label{2asol}
\ee
%%%%%
where we have the novel feature of the $F_\4$ term given in terms of
the constant 3-form $\Phi_3$ on the flat $\R^8$ transverse space.
Using the results for the Ricci tensor presented in (\ref{riccicomp}),
we find after straightforward algebra that the type IIA
equations of motion are satisfied if $H$ and $K$ satisfy
%%%%%
\be
\square \, H = 0\,,\qquad \square \, K = -\ft16 \mu^2\, |\Phi_3|^2\,,
\ee
%%%%%
where $\square =\del_i\del_i$ is the Laplacian in the flat transverse 
space.  As in the type IIB example, we can take 
%%%%%
\be
K= c_0 -\sum_i \mu_i^2\, z_i^2\,, \qquad H= 1 + \fft{Q}{r^6}\,,
\ee
%%%%%
where $\sum_i \mu_i^2 = \ft1{12}\, \mu^2\, |\Phi_\3|^2$, and $r^2=z_i^2$.

   Note that we can lift this intersecting solution back to $D=11$
supergravity, by using the standard Kaluza-Klein reduction
%%%%%
\bea
d\hat s_{11}^2 &=& e^{-\fft16\phi}\, ds_{10}^2 + e^{\fft43\phi}\, 
(dz_9+ \cA_1)^2\,,\nn\\
\hat A_\3 &=& A_\3 + (dy+\cA_\1)\wedge A_\2\,.
\eea
%%%%%
Since we have $\cA_\1=0$ here, we find that the lifting just gives
%%%%%
\bea
d\hat s_{11}^2 &=& H^{-2/3}\, (2 dx^+\, dx^- + K\, {dx^+}^2 + dz_{10}^2) +
         H^{1/3}\, dz_i^2 \,,\nn\\
\hat F_\4 &=& \mu\, dx^+\wedge \Phi_\3 + dy\wedge dH^{-1}\wedge dx^+\wedge 
dx^-\,.\label{d11solution}
\eea
%%%%%
(Note that $z_i$ has $1\le i\le 8$, denoting the transverse coordinates
in the 8-dimensional transverse space, while $z_{10}$ denotes the 
extra coordinate of eleven dimensions.)
This describes a membrane living in a pp-wave background in eleven 
dimensions.

\subsection{Supersymmetry of the type IIA string on a pp-wave}
\label{2astringwave}

   This can be studied most simply by lifting the solution to
eleven-dimensional supergravity.  In $D=11$, we choose the natural 
vielbein basis for the metric in (\ref{d11solution}), giving
%%%%%
\bea
\hat e^{10} &=& H^{-1/3}\, dz_{10}\,,\nn\\
\hat e^i &=& H^{1/24}\, e^i = H^{1/6}\, dz^i\,,\qquad 1\le i\le 8\,,\nn\\
\hat e^+ &=& H^{1/24}\, e^+ = H^{-1/3}\, dx^+\,,\label{hatframe}\\
\hat e^- &=& H^{1/24}\, e^- = H^{-1/3}\, (dx^+ + \ft12 K\, dx^-)\,.\nn
\eea
%%%%%
Using standard Kaluza-Klein formulae, we can express the spin connection 
$\hat \omega^{AB}$ in $D=11$ in terms of $D=10$ quantities, and hence we 
find that $\hat\nabla\equiv d+\ft14 \hat\omega^{AB}\, \Gamma_{AB}$ is
given by
%%%%%
\be
\hat\nabla= d - \ft1{12} H^{-7/6}\, H_i\, (2\Gamma_{+i}\, \hat e^+
 + 2\Gamma_{-i}\, \hat e^- + \Gamma_{ij}\, \hat e^j) + \ft14 H^{-1/6}\, 
K_i\, \Gamma_{-i}\, \hat e^+ + H^{-7/6}\, \Gamma_i\, 
\Gamma_{\underline{10}}\, 
\hat e^{10}\,,
\ee
%%%%%
where we are denoting by $\Gamma_{\underline{10}}$ the
eleven-dimensional Dirac matrix associated with the extra coordinate
$z_{10}$ of the M-theory circle.  (We underline the ``10'' to avoid
the danger of confusion between $\Gamma_{10}$ and
$\Gamma_1\, \Gamma_0$.)  We reserve the subscript ``9'' on a Dirac matrix
for the spatial direction associated with the $x^{\pm} \equiv (x^9\pm
x^0)/\sqrt2$ coordinates.

   The gravitino transformation rule is given by $\delta\hat\psi_A =
\hat\nabla_A\, \ep + \hat T_A\, \ep$, where
%%%%
\be
\hat T_A = -\ft1{288}\, \Gamma_A{}^{BCDE}\, \hat F_{BCDE} + \ft1{36}
\hat F_{ABCD}\, \Gamma^{BCD}\,.
\ee
%%%%%
The $D=11$ frame components of the field strength $\hat F_\4$ can be
read off from (\ref{d11solution}) and (\ref{hatframe}), giving
%%%%%
\be
\hat F_{+ijk} = \mu\, H^{-1/6}\, \Phi_{ijk}\,,\qquad 
\hat F_{i\underline{10}+-} = H^{-7/6}\, H_i\,.
\ee
%%%%%
(Again, we underline the ``10'' subscript here to avoid potential
confusion with subscripts ``1'' followed by ``0.''

    Substituting into the $D=11$ supersymmetry transformation rule,
we find that $\delta\hat\psi_A=0$ implies
%%%%%
\bea
&&\del_{10}\, \ep +\ft{\im}{12}\, \mu \, H^{-1/2}\, W\, 
\Gamma_{\underline{10},-}\, \ep
+ \ft16 H_i\, \Gamma_i\, (\Gamma_{\underline{10}}- \Gamma_{-+})\, \ep=0\,\nn\\
&& \del_i\,\ep + \ft{\im}{24} \mu\, (\Gamma_i\, W + 3 W\, \Gamma_i)
\Gamma_-\, \ep
 + \ft1{12} H^{-1}\, [2H_i\, \Gamma_{\underline{10}}\, \Gamma_{-+} 
+ H_j\, 
\Gamma_{ij}\, (1-\Gamma_{\underline{10}}\Gamma_{-+})]\, \ep=0\,,\nn\\
&&\del_-\,\ep + \ft16 H^{-3/2}\, H_i\, \Gamma_i\, \Gamma_-\, 
(1+\Gamma_{\underline{10}})\, \ep
=0\,,\\
&& \del_+\, \ep -\ft12 K\, \del_-\, \ep \nn\\
&&+ \ft14 H^{-1/2}\, [
K_i\, \Gamma_{-i} - \ft{\im}{3}\mu\, W\, 
(1+\Gamma_-\, \Gamma_+)]\, \ep 
+\ft16 H^{-3/2}\, H_i\, \Gamma_i\, \Gamma_+\, 
(1-\Gamma_{\underline{10}})\, \ep=0\,,\nn
\eea
%%%%%
where we have defined 
%%%%%
\be
W\equiv \ft{\im}{6}\, \Phi_{ijk}\, \Gamma_{ijk}\,.
\ee
%%%%%
The four lines come from $\delta\hat\psi_{10}=0$, $\delta\hat\psi_i=0$, 
$\delta\hat\psi_-=0$, $\delta\hat\psi_+=0$ respectively.  Note that
indices on $\del$ are all coordinate indices, while those on the Dirac
matrices are all tangent-frame indices.

   It is easy to see from the equations above that we must have
%%%%%
\bea
&&\Gamma_-\, (1+\Gamma_{\underline{10}})\, \ep=0\,,\qquad 
            \Gamma_+\, (1-\Gamma_{\underline{10}})\, \ep=0\,,\nn\\
&&\del_{10}\,\ep=0\,,\qquad \del_-\, \ep=0\,,\qquad \del_+\, \ep=0\,,
\label{stringcon}
\eea
%%%%%
and that if we define $\ep = H^{-1/6}\, \eta$, then $\eta(z_i)$ must 
satisfy
%%%%%
\be
\del_i\, \eta + \ft{\im}{24} \mu\, (\Gamma_i\, W + 3 W\, \Gamma_i)\, 
\Gamma_-\, 
\eta=0\,,\quad
K_i\, \Gamma_{-i}\, \eta - \ft{\im}{3}\mu\, W\, (1+\Gamma_-\, \Gamma_+)\, 
\eta=0\,.\label{wavecon}
\ee
%%%%%
The first line in (\ref{stringcon}) represents the usual projections
that one encounters for a string solution, implying a half-breaking of
supersymmetry.  The conditions in (\ref{wavecon}) are the same as
those encountered in \cite{supernum} for pure pp-waves, except that here
we have the added requirement of $x^+$ independence.  As described in 
\cite{supernum}, the first equation in (\ref{wavecon}) implies that 
%%%%%
\be
\eta = \Big(1-\ft{\im}{24}\, \mu\, (\Gamma_i\, W + 3 W\, \Gamma_i)\,
\Gamma_-\Big) \, \chi\,,
\ee
%%%%%
where $\chi$ is independent of the $z_i$.  Since here we also have
$\del_+\ep=0$, it follows, using arguments analogous to those in
\cite{supernum}, that $\chi$ must satisfy
%%%%%
\be
W\, \chi=0\,,\qquad (\mu^2\, z_i\, W^2+ 8 K_i)\, \Gamma_i\, \chi=0\,. 
\label{chicon}
\ee
%%%%%

   Following \cite{clps}, we shall define the chirality operator
$\Gamma_{11} \equiv -\Gamma_{\underline{10}}$, where it will be
recalled that $\Gamma_{\underline{10}}$ denotes the eleven-dimensional
Dirac matrix in the direction of the M-theory circle.  Multiplying
the expressions on the first line in (\ref{stringcon}) by $\Gamma_9$,
they can be written as 
%%%%%
\be
(1+\Gamma_{09})(1-\Gamma_{09}\, \Gamma_*)\,\ep=0\,,\qquad
(1-\Gamma_{09})(1+\Gamma_{09}\, \Gamma_*)\,\ep=0\,,
\ee
%%%%%
where $\Gamma_*\equiv \Gamma_1\cdots \Gamma_8$ as in (\ref{gammastar}).  
Adding and subtracting these equations, we find that the conditions on 
the first line in (\ref{stringcon}) imply and are implied by
%%%%%
\be
(1-\Gamma_*)\, \ep=0\,.
\ee
%%%%%
Thus we see that the string component of the string/pp-wave intersection
in the type IIA theory imposes precisely the gamma-matrix projection
condition that we used in the physical-gauge fixing in section 
\ref{2aphysgauge}.

\subsection{Strings on pp-waves with odd numbers of Killing spinors}

      Here, we study the number of Killing spinors that can occur in
the type IIA and type IIB strings in pp-wave backgrounds. As we stated
earlier, the type IIA and type IIB cases are related by T-duality.
Thus it suffices to study the type IIA example.  We take the constant
3-form to be \cite{clppp,supernum}
%%%%%
\bea
\phi_\3&=&m_1\, dz^{123} + m_2\, dz^{145} + m_3\, dz^{167} +
m_4\, dz^{246} \nn\\
&& + m_5\, dz^{257} + m_6\, dz^{347} + m_7\, dz^{356}
\,.
\eea
%%%%%%
As we saw in (\ref{2astringwave}), for both the standard and supernumerary
Killing spinors we must have $W\,\chi=0$, which implies that
%%%%%
\be
m_1 + m_2 + m_3 - m_4 + m_5 + m_6 + m_7=0\,.
\label{zeroeigenvalue}
\ee
%%%%%
The standard Killing spinors satisfy the three conditions
%%%%
\be
\Gamma_-\, \chi=0\,,\qquad W\, \chi=0\,,\qquad 
(1-\Gamma_{\underline{10}})\, \chi=0\,.
\ee
%%%%%
Since $\Gamma_-$, $\Gamma_{\underline{10}}$ and $W$ commute, it
follows that the total number of Killing spinors is a quarter of
the number of spinors annihilated by $W$.   

   In the case of pure pp-waves with no string present, the number of
standard Killing spinors is 16, and is always independent of the
structure of $W$.  The detailed structure of $W$ determines their
$x^+$ dependence, with only those that are annihilated by $W$ being
independent of $x^+$.  In the present case, where there is a string on
the pp-wave background, we saw in section \ref{2astringwave} that the
Killing spinors must be $x^+$-independent, and so it follows that even
the number of standard Killing spinors will now depend on the
structure of $W$.  It is straightforward to see that a generic
set of $m_i$ satisfying the condition (\ref{zeroeigenvalue}) will give
rise to one standard Killing spinor, with a maximum of four standard
Killing spinors being achievable for suitable special choices of the
$m_i$.

       The conditions for supernumerary Killing spinors are slightly
more restrictive; they are given by
%%%%%%
\bea
\Gamma_+\, \chi=0\,,\qquad W\, \chi=0\,,\qquad
(1 + \Gamma_{\underline{10}})\, \chi=0\,,\qquad
(\mu^2\, z_i\, W^2 + 8 K_i)\, \Gamma_i\, \chi=0\,.
\eea
%%%%%
For a generic set of $m_i$ satisfying (\ref{zeroeigenvalue}),
there is one supernumerary Killing spinor, with a maximum of four for special
values of $m_i$.

   Let us look at some examples in detail.  If there is only one
non-vanishing $m_i$, $W$ has no zero eigenvalues, and so there are neither
standard nor supernumerary Killing spinors.  

   For two non-vanishing $m_i$, say $m_1$ and $m_2$, we can obtain
Killing spinors if we choose $m_1+m_2=0$.  There are then four standard
Killing spinors and four supernumerary Killing spinors.

   For three non-vanishing $m_i$, say $m_1$, $m_2$ and $m_3$, there
exist Killing spinors if $m_1+m_2+m_3=0$.  For generic such choices, 
there are two standard Killing spinors and two supernumerary Killing
spinors.  

   In general, when there are four or more non-vanishing $m_i$, there
will be one standard Killing spinor and one supernumerary Killing spinor
provided that the $m_i$ satisfy (\ref{zeroeigenvalue}).  However, more
Killing spinors can arise in special cases.  For example, if we have
$m_1=1$, $m_2=-1$, $m_3=1$ and $m_4=1$, we have three standard Killing
spinors and one supernumerary Killing spinor.  For a more complicated
example, $m_1=1$, $m_2=-1$, $m_3=1$, $m_4=1$, $m_5=1$, $m_6=-1$, we
have four standard Killing spinors and one supernumerary Killing spinor,
giving a total of five Killing spinors.  If $m_1=2$, $m_2=-2$, $m_3=1$,
$m_4=1$, $m_5=1$, $m_6=-1$, we have two standard Killing spinors and one
supernumerary Killing spinor, giving a total of three.  These latter two
examples are of interest since the total number of Killing spinors is
odd.

\subsection{Further solutions for strings in pp-waves}
\label{another2b}

\noindent{\bf A type IIB generalisation}

   For this example, we again take the metric and dilaton to be
%%%%%
\bea
ds_{10}^2 &=& H^{-3/4}\, (2 dx^+\, dx^- + K\, {dx^+}^2) + H^{1/4}\, 
    dz_i^2\,,\nn\\
\phi &=& -\ft12 \log H\,,
\eea
%%%%%
but now the solution is supported by the NS-NS and R-R 3-form field
strengths $G_\3$ and $F_\3$:
%%%%%
\be
G_\3 = dH^{-1}\wedge dx^+\wedge dx^-\,,\qquad F_\3 = \mu\, dx^+ \wedge
 \Phi_\2\,,
\ee
%%%%%
where $\Phi_\2$ is a constant 2-form in the eight-dimensional transverse 
space.
It can be verified that this solves the type IIB equations of motion
provided that $H$ and $K$ satisfy
%%%%%
\be
\square H=0\,,\qquad \square K = -\ft12 \mu^2\, |\Phi_\2|^2
\ee
%%%%%
in the transverse space.

\noindent{\bf A type IIA generalisation}

   We can generalise the type IIA string/pp-wave given in (\ref{2asol})
by adding in a constant R-R 2-form field strength term, so that we
now have
%%%%%
\bea
ds_{10}^2 &=& H^{-3/4}\, (2 dx^+\, dx^- + K\, {dx^+}^2) + H^{1/4}\, 
    dz_i^2\,,\qquad \phi=-\ft12 \log H\,,\nn\\
F_\4 &=& \mu\, dx^+ \wedge \Phi_\3\,,\qquad F_\3= dH^{-1}\wedge dx^+\wedge 
          dx^-\,,\qquad F_2 = \td \mu\, dx^+\wedge \Phi_\1\,.
\label{2asol2}
\eea
%%%%%   
Substituting into the type IIA equations of motion, we find that
they are satisfied provided that $H$ and $K$ satisfy
%%%%%
\be
\square\, H=0\,,\qquad \square \, K= -\ft16 \mu^2\, |\Phi_\3|^2 - \td\mu^2\, 
|\Phi_\1|^2\,.
\ee
%%%%%

    Some similar solutions, corresponding to cases of D-branes in
pp-wave backgrounds that originate from AdS$_3\times S^3\times R^4$,
can be found in \cite{kumar1,kumar2}.

\section{Conclusions}

    In this paper, we have studied the relation between the occurrence
of Killing spinors in a pp-wave background, and the worldsheet
supersymmetry of the associated string action after gauge fixing.  In
the light-cone gauge, the string action gives a free theory with masses
for the bosonic and fermionic fields \cite{bermalnas}. 

   The Killing spinors in the pp-wave background comprise 16
``standard'' Killing spinors, which are present in arbitrary pp-wave
backgrounds, plus an additional $N_{\rm sup}$ supernumerary Killing
spinors, with $0\le N_{\rm sup}\le 16$, which arise only in special
cases.  We showed that in the light-cone gauge, none of the 16
standard Killing spinors give rise to any linearly-realised worldsheet
supersymmetries. This is because the projection condition for the
linearly-realised worldsheet supersymmetries arising from the local
kappa symmetry and target-spacetime supersymmetry after light-cone
gauge fixing is orthogonal to the projection condition satisfied by
the standard Killing spinors.  By contrast, the supernumerary Killing
spinors satisfy the opposite projection condition, and so these are
precisely the ones that are associated with linearly-realised
worldsheet supersymmetries in the light-cone gauge.
 
   Depending upon the specific details of a pp-wave solution that
admits supernumerary Killing spinors, these Killing spinors may either
depend upon the coordinate $x^+$, or else be independent of $x^+$.
Since $x^+$ is set equal to $\tau$ in the light-cone gauge, the
question of $x^+$ dependence determines whether or not the associated
linearly-realised worldsheet supersymmetries commute with the
Hamiltonian.  If they do commute (\ie when the Killing spinors do not
depend upon $x^+$), then the linear supersymmetry ensures that the bosonic
and fermionic coordinates will have paired sets of equal mass terms.
If the supernumerary Killing spinors instead depend upon $x^+$, then
even though there are still linearly-realised worldsheet
supersymmetries, they no longer imply any equality of the boson and
fermion masses.

   Since the light-cone gauge provides the standard passage between
the Green-Schwarz and the Neveu-Schwarz-Ramond formalisms, the
considerations of this paper could lead to a better understanding of
the NSR formalism in the presence of Ramond-Ramond backgrounds, such
as the constant RR forms that we have discussed. The passage to the
NSR formalism endows a specific worldsheet (1,1) supersymmetry with a
special importance, being the rigid gauge-fixed remnant of the NSR
(1,1) local worldsheet supersymmetry. This passage makes use of
Killing spinors on the spacetime background for the light-cone gauge
fixed theory in order to change the fermionic worldsheet variables
from spinor to vector representation of the transverse-space structure
group ({\it c.f., e.g.}\ \cite{gsnsr}). \footnote{An alternative approach
is to employ the bosonisation procedure for the worldsheet fermions. 
Bosonisation provides a direct map between the worldsheet fields in
the NSR formalism, where they transform as space-time vectors, and the
worldsheet fields in the GS formalism, where they
transform as space-time spinors, {\it c.f., e.g.}\ \cite{peskin}.}

 Additional NSR worldsheet
supersymmetries beyond the original (1,1) arise depending on the
geometry of the background. For example, if the transverse geometry is
K\"ahler, one obtains (2,2) supersymmetry; if it is hyper-K\"ahler,
one obtains (4,4) supersymmetry, {\it etc.} These additional NSR
supersymmetries are distinguished from the initial (1,1) supersymmetry
in that they all involve complex structures on the target manifold of
the resulting NSR worldsheet sigma model. Thus, the linearly realised
and $x^+$ independent lightcone gauge supersymmetries that we have
discussed in this paper can be expected to translate to rigid NSR
supersymmetries for superstrings propagating on the class of pp-wave
backgrounds that we have discussed.

   In this paper, we have found that a quite different situation
arises if one chooses the physical gauge instead of the light-cone
gauge.  We showed that in this case the projection conditions for the
residual worldsheet supersymmetries that arise from the original kappa
symmetry and target-spacetime supersymmetry now imply that the
linearly-realised worldsheet supersymmetries come both from standard
and from supernumerary Killing spinors.  We also showed that,
consistently with this, there are in fact no mass terms at all in the
string worldsheet action in the physical gauge, either for the bosonic
or the fermionic coordinates.  This emphasises the fact that the mass
terms in the string worldsheet actions in the light-cone gauge can be
viewed as artefacts of the specific gauge-fixing procedure.

   We also obtained new supergravity solutions that describe strings
in pp-wave backgrounds. These can be viewed as the classical solitonic
realisations of string actions whose target spacetimes are pp-waves.
The supergravity solutions naturally describe these string actions in
the physical gauge, since the $x^\pm$ coordinates of the pp-wave are
the same as the worldsheet coordinates of the string component of the
classical solution.  We find that indeed the supersymmetries of these
classical supergravity solutions coincide with the supersymmetries
that one finds for the string actions in the corresponding pp-wave
background, upon imposition of the physical gauge conditions.  In
particular, we find that there can be odd numbers of Killing spinors
in these solutions.

\section*{Acknowledgments}

K.S.S. would like to thank Chris Hull for discussions.  M.C.,
C.N.P. and K.S.S. thank the Isaac Newton Institute and CERN,
C.N.P. and K.S.S. thank the Ecole Normale, Paris,  M.C. thanks the  New
Center for Theoretical Physics at Rutgers University and K.S.S. thanks
the Institut des Hautes Etudes Scientifiques, for hospitality and
support at various times during the course of this work.


\begin{thebibliography}{99}


\bibitem{kg} J. Kowalski-Glikman,
{\it Vacuum States In Supersymmetric Kaluza-Klein Theory,}
Phys. Lett. {\bf B134}, 194 (1984).
%%CITATION = PHLTA,B134,194;%% 

\bm{blafighulpap} M. Blau, J. Figueroa-O'Farrill, C. Hull and
G. Papadopoulos, {\it A new maximally supersymmetric background of IIB
superstring theory}, JHEP {\bf 0201} (2002) 047, hep-th/0110242.
%%CITATION = HEP-TH 0110242;%%

\bibitem{blafighulpap2} M. Blau, J. Figueroa-O'Farrill, C. Hull and
G. Papadopoulos, {\it Penrose limits and maximal supersymmetry},
hep-th/0201081.
%%CITATION = HEP-TH 0201081;%%

\bm{penrose} R. Penrose, {\it Any space-time has a plane wave as a limit},
in {\sl Differential geometry and relativity}, Reidel, Dordrecht, 1976.


\bm{met} R.R. Metsaev,{\it Type IIB Green-Schwarz superstring in plane
wave Ramond-Ramond background}, Nucl. Phys. {\bf B625}, 70 (2002),
hep-th/0112044.
%%CITATION = HEP-TH 0112044;%%

\bm{bermalnas} D. Berenstein, J. Maldacena and H. Nastase, {\it
Strings in flat space and pp waves from N = 4 super Yang Mills},
hep-th/0202021. 
%%CITATION = HEP-TH 0202021;%% 

\bibitem{clppp} M. Cveti\v c, H. L\"u and C.N. Pope,
{\it Penrose limits, pp-waves and deformed M2-branes,} hep-th/0203082.
%%CITATION = HEP-TH 0203082;%% 

\bm{supernum}
M. Cveti\v c, H. L\"u and C.N. Pope,
{\it M-theory pp-waves, Penrose limits and supernumerary supersymmetries,}
hep-th/0203229, to appear in Nucl. Phys. {\bf B}.
%%CITATION = HEP-TH 0203229;%%

\bm{gahu} J. Gauntlett and C.M. Hull, 
{\it pp-waves in 11-dimensions with extra supersymmetry,}
JHEP {\bf 0206}, 013 (2002), hep-th/0203255.
%%CITATION = HEP-TH 0203255;%%

\bm{lupo} H. L\"u and J.F. Vazquez-Poritz,
{\it Penrose limits of non-standard brane intersections,}
Class.\ Quant.\ Grav.\  {\bf 19}, 4059 (2002), hep-th/0204001.
%%CITATION = HEP-TH 0204001;%%

\bibitem{clps} M. Cveti\v c, H. L\"u, C.N. Pope and K.S. Stelle,
{\it T-duality in the Green-Schwarz formalism, and the
massless/massive IIA  duality map,}
Nucl.\ Phys.\ {\bf B573}, 149 (2000),
hep-th/9907202.
%%CITATION = HEP-TH 9907202;%%


\bm{dwpp} B. de Wit, K. Peeters and J. Plefka,
{\sl Superspace geometry for supermembrane backgrounds},
Nucl. Phys. {\bf B532} (1998) 99, hep-th/9803209.

\bm{duhoinst} M.J. Duff, P.S. Howe, T. Inami and K.S. Stelle,
{\it Superstrings in $D = 10$ from supermembranes in $D = 11$},
Phys. Lett. {\bf 191B} (1987) 70.

\bm{cohakewa}
R. Corrado, N. Halmagyi, K.D. Kennaway and N.P. Warner,
{\it Penrose limits of RG fixed points and pp-waves with background fluxes,}
hep-th/0205314.
%%CITATION = HEP-TH 0205314;%%

\bm{kumar1} A. Kumar, R.R. Nayak and Sanjay,
{\it D-brane solutions in pp-wave background,}
Phys.\ Lett. {\bf B541}, 183 (2002), hep-th/0204025.
%%CITATION = HEP-TH 0204025;%%

\bm{kumar2} A. Biswas, A. Kumar and K.L. Panigrahi,
{\it $p - p'$ branes in pp-wave background,} hep-th/0208042.
%%CITATION = HEP-TH 0208042;%%

\bm{gsnsr} 
C.M. Hull, {\it Compactifications of the heterotic superstring,} 
Phys. Lett. {\bf
B178}, 357 (1986);\newline  M.D. Freeman, C.N. Pope, C.M. Hull and 
K.S. Stelle,
{\it Space-time versus world sheet supersymmetry 
in the heterotic string,} Phys.\
Lett.\ {\bf B185}, 351 (1987).

\bm{peskin}
M.E. Peskin,
{\it Introduction To String And Superstring Theory. 2,}
SLAC-PUB-4251,
{\it Lectures presented at the 1986 Theoretical Advanced Study Institute
in Particle Physics, Santa Cruz, Calif., Jun 23 - Jul 19, 1986}.


\end{thebibliography}
\end{document}